\documentclass[twocolumn,aps,prd,showpacs,nofootinbib,superscriptaddress,
preprintnumbers,floatfix,10pt]{revtex4-1}

\usepackage{amsmath}
\usepackage{amsfonts}
\usepackage{amssymb}
\usepackage{epsfig}
\usepackage{graphicx}
\usepackage{color}
\usepackage{slashed}
\usepackage{epstopdf}
\usepackage{dsfont}
\usepackage{bbold}
\usepackage{mathtools}
\usepackage{array}
\usepackage{verbatim}

\newcommand{\PreserveBackslash}[1]{\let\temp=\\#1\let\\=\temp}
\newcolumntype{C}[1]{>{\PreserveBackslash\centering}p{#1}}

\usepackage{tikz}

\usetikzlibrary{positioning,decorations.pathmorphing,fit}

\def\xr{1.9}
\def\yr{0.6}

\newcommand\addvmargin[1]{
  \node[fit=(current bounding box),inner ysep=#1,inner xsep=0]{};
}

\usepackage{hyperref}

\newcommand{\STr}{\text{STr}}
\newcommand{\Tr}{\text{Tr}}
\newcommand{\Eqref}[1]{Eq.~\eqref{#1}}

\newcommand{\Nf}{N_{\mathrm{f}}}
\newcommand{\pat}{\partial_t}

\DeclareMathOperator{\U}{U}
\DeclareMathOperator{\g}{\gamma} 
\newcommand{\barh}{\bar{h}}

\begin{document}

\preprint{}

\title{Mean-field theory for self-interacting relativistic Luttinger fermions} 

\author{Holger Gies}
\email{holger.gies@uni-jena.de}
\affiliation{Theoretisch-Physikalisches Institut, 
Abbe Center of Photonics, Friedrich Schiller University Jena, Max Wien 
Platz 1, 07743 Jena, Germany}
\affiliation{Helmholtz-Institut Jena, Fr\"obelstieg 3, D-07743 Jena, Germany}
\affiliation{GSI Helmholtzzentrum für Schwerionenforschung, Planckstr. 1, 
64291 Darmstadt, Germany} 
\author{Marta Picciau}
\email{marta.picciau@uni-jena.de}
\affiliation{Theoretisch-Physikalisches Institut, 
Abbe Center of Photonics, Friedrich Schiller University Jena, Max Wien 
Platz 1, 07743 Jena, Germany}

\begin{abstract}
We investigate a class of quantum field theories with relativistic Luttinger
fermions and local self-interaction in scalar channels. For an understanding of
possible low-energy phases, we first classify the set of mass terms arising from
scalar fermion bilinears. For large flavor numbers, we show that each of our models features a coupling
branch in which the theory is asymptotically free. In order to address the
long-range behavior, we use mean-field theory which is exact in the limit of
large flavor numbers. We identify two models which undergo dimensional
transmutation, interconnecting the asymptotically free high-energy regime with
an ordered low-energy phase sustaining a vacuum condensate. We also study the
analytic structure of the Luttinger-fermionic propagator in the various possible
gapped phases.

\end{abstract}


\maketitle


\section{Introduction}
\label{intro}

Luttinger fermions are effective degrees of freedom of non-relativistic solid
state physics \cite{LuttingerPhysRev.102.1030,Abrikosov:1974a} used to describe,
e.g., materials with quadratic band touching/crossing points involving general
spin-orbit couplings  \cite{Murakami:2004zz,Moon:2012rx,Savary:2014gka}. These
systems can feature a rich set of quantum critical phenomena
\cite{Herbut:2014lfa,Janssen:2015xga,Janssen:2015jga,Boettcher:2016wft,
Janssen:2016xvc,Boettcher:2016iiv, Ray:2018gtp,Ray:2020mlg,Ray:2021moi,
Ray:PhD,Dey:2022lkx}. Inspired by the diverse set of structures emerging from
such long-range degrees of freedom, the generalization of Luttinger fermions to
fundamental degrees of freedom of relativistic quantum field theories has
recently been studied \cite{Gies:2023cnd}.

Since the resulting relativistic Luttinger operator is quadratic, the mass
dimension of the field agrees with that of standard scalar fields which allows
for the construction of a large number of perturbatively renormalizable quantum
field theories in 3+1 dimensional spacetime. Specifically, self-interacting
theories of Luttinger fermions are renormalizable and can also be asymptotically
free \cite{Gies:2023cnd}. As a consequence, such quantum field theories can
serve as a novel building block for high-energy complete theories for particle
physics.

Another unorthodox feature of these theories becomes visible in the pole
structure of the propagator where properties familiar from those of a
higher-derivative theory
\cite{Pais:1950za,Lee:1970iw,Stelle:1976gc,Grinstein:2007mp} can be found
\cite{Gies:2023cnd}. For a standard mass term, not only the standard particle
pole but also a tachyonic pole appears. The latter comes with a negative
residue, characterizing a so-called ghost. Naively, this is often taken as an
indication of non-unitarity interpreted as a consequence of Ostrogradsky's
theorem \cite{Ostrogradsky:1850fid}, even though many different viewpoints on
such ghost states exist in the literature, see e.g.,
\cite{Lee:1970iw,Narnhofer:1978sw,Hawking:2001yt,Bender:2007wu,Grinstein:2007mp,%
Garriga:2012pk,Salvio:2014soa,Smilga:2017arl,Becker:2017tcx,Anselmi:2018kgz,%
Gross:2020tph,Donoghue:2021eto,Platania:2019qvo,Deffayet:2023wdg}.

In the present work, we concentrate on a set of simple example theories
involving self-interacting relativistic Luttinger fermions. More specifically,
we concentrate on  massless classical actions with local scalar or pseudo-scalar
interactions. In addition to an investigation of the high-energy behavior
characterized by the beta functions of the couplings, we explore the long-range
behavior of these theories using the mean-field approximation as a simple tool,
being exact in the limit of large flavor number $\Nf$. We pay specific attention
to the possible condensates and the long-range phase diagrams.
Specifically, we identify two models that feature asymptotic freedom in
the ultraviolet (UV), undergo dimensional transmutation in the sense of Coleman
and Weinberg \cite{Coleman:1973jx}, and exhibit condensate formation in the
long-range limit.

Since we expect the long-range phases to be characterized by a massive spectrum
we start our exploration with a classification of possible mass terms for the
relativistic Luttinger fermions. As the relativistic Abrikosov algebra needs to
be spanned by a reducible representation of the underlying Clifford algebra,
there is a larger set of possible mass terms. The latter is reminiscent to mass
terms of relativistic 2+1 dimensional Dirac materials where several mass terms
can describe different patterns of gap formation
\cite{Semenoff:1984dq,Appelquist:1986fd,Haldane:1988zza,Kane:2004bvs}. 

Interestingly, the richer set of mass terms also goes along with a more
intricate analytic structure of the corresponding propagators. We observe that
the two asymptotically free models with low-energy condensate formation at the
mean-field level do not feature tachyonic mass poles but a complex pair of poles
or a branch cut.

Our paper is organized as follows: We begin in Sect.~\ref{sec:Lfermions} with a
short summary of relativistic Luttinger fermions following \cite{Gies:2023cnd}.
In Sect.~\ref{sec:massterms}, we present a set of different mass terms for
relativistic Luttinger fermions. Section~\ref{sec:selfintmodels} introduces the
set of models discussed in the present work. Here, we verify that each one
features an asymptotically free coupling branch by computing the perturbative
one-loop beta function. In Sect.\ref{sec:MFT}, we solve each model in a
mean-field approximation exploring their potential for condensate formation.   
Section \ref{sec:propagators}, we study the analytic structure of the gapped
Luttinger propagators in the complex momentum square plane. Conclusions are
given in Sect.~\ref{sec:conc}.  

\section{Relativistic Luttinger fermions}
\label{sec:Lfermions}

We define field theories of relativistic Luttinger fermions in terms of their
classical action. Focusing on 4-dimensional spacetime, the action of the free
theory reads \cite{Gies:2023cnd},
\begin{equation}
 S=\int d^4 x \left[ \bar\psi G_{\mu\nu} (i\partial^\mu) (i \partial^\nu) \psi 
\right],
\label{eq:Sfree}
\end{equation}
where $\psi$ denotes a spinor with $d_\gamma$ components. Correspondingly,
$G_{\mu\nu}$ represents a set of $d_\gamma\times d_\gamma$ matrices
labeled by the Lorentz indices $\mu,\nu=0, \dots, 3$. These matrices satisfy
the relativistic version of the Abrikosov algebra
\cite{Abrikosov:1974a,Janssen:2015xga,Gies:2023cnd}
\begin{equation}
    \{G_{\mu\nu}, G_{\kappa\lambda}\} = - \frac{2}{3} 
   g_{\mu\nu}g_{\kappa\lambda}+\frac{4}{3} (g_{\mu\kappa} g_{\nu\lambda} + 
   g_{\mu\lambda}g_{\nu\kappa}),
   \label{eq:AbrikosovA}
\end{equation} 
where the right-hand side involves the Minkowski metric $g=\text{diag}(+,-,-,-)$
and is also implicitly understood to be proportional to the identity
$\mathbb{1}_{d_\gamma}$ in spinor space. With respect to the Lorentz indices,
the matrices $G_{\mu\nu}$ are symmetric, $G_{\mu\nu}=G_{\nu\mu}$, and traceless,
$G^\mu{}_\mu=g^{\mu\nu} G_{\mu\nu}=0$, implying that 9 linearly independent
elements are needed to span the Abrikosov algebra \eqref{eq:AbrikosovA}. With
respect to the spin indices, we can choose $G_{0i}$ anti-hermitean whereas
$G_{00}$ and $G_{ij}$ can be chosen hermitean for all $i,j=1,2,3$.  

Finally, the conjugate spinor in \Eqref{eq:Sfree} is defined by
$\bar\psi=\psi^\dagger h$ involving the spin metric $h$. Choosing $h$ hermitean
$h^\dagger=h$, the requirement that the classical action is real,
$S\in\mathbb{R}$, imposes the conditions
\begin{equation}
    \{h,G_{0i}\}=0, \quad [h,G_{ij}]=0, \quad 
   [h,G_{\underline{\mu}\underline{\mu}}]=0,
   \label{eq:hcond}
\end{equation}
where underscored indices are exempted from Einstein's sum convention.

Both sets of algebraic conditions \eqref{eq:AbrikosovA} and \eqref{eq:hcond} can
be spanned by a Euclidean Dirac algebra,
\begin{equation}
    \{\gamma_A,\gamma_B\} = 2 \delta_{AB},
    \label{eq:EuclidDirac} 
\end{equation}
with $d_e$ hermitean elements, $A,B=1,2, \dots, d_e$. Whereas the
irreducible representation of the Abrikosov algebra would, in principle, require
only $d_e=9$, the additional reality conditions \eqref{eq:hcond} demands for
$d_e=11$. The latter implies that $d_\gamma=2^{\lfloor
d_{\text{e}}/2\rfloor}=32$ characterizes the irreducible representation of
relativistic Luttinger fermions. An explicit representation of the $G_{\mu\nu}$
in terms of the Euclidean Dirac matrices $\gamma_A$ is given in
App.~\ref{sec:AppA}.  Setting, for instance, $G_{0i}=i \sqrt{\frac{2}{3}}
\gamma_{A=i}$, all other $G_{\mu\nu}$ are real linear combinations of
$\gamma_{A=4,\dots,9}$, and the spin metric can be chosen as 
\begin{equation}
    h=\gamma_1 \gamma_2 \gamma_3 \gamma_{10}.
    \label{eq:spinmetric}
\end{equation}
The free field equation for Luttinger fermions derived from \Eqref{eq:Sfree}
reads
\begin{equation}
    G_{\mu\nu} \partial^\mu \partial^\nu \psi =0.
\end{equation}
Using the Abrikosov algebra, it follows straightforwardly that the Luttinger
operator squares to (the square of) the Klein-Gordon operator, $(G_{\mu\nu}
\partial^\mu \partial^\nu)^2= (\partial^2)^2$, implying that each of the 32
components of $\psi$ satisfies a relativistic wave equation. 

\section{Mass terms}
\label{sec:massterms}

In order to classify different possibilities of gap formation potentially
occurring in self-interacting models studied below, let us first investigate the
different mass terms that can be constructed for Luttinger fermions. As basic
requirements, we are interested in Lorentz invariant bilinear and real terms
that we can add to the action. 

For this, let us first recall that the Abrikosov algebra is separately invariant
under Lorentz transformations
\begin{equation}
    G_{\mu\nu}\to G_{\kappa\lambda} \Lambda^\kappa{}_\mu \Lambda^\lambda{}_\nu,
    \label{eq:GLtrafo}
\end{equation}
where $\Lambda^\kappa{}{}_\mu$ is the transformation matrix of Lorentz tensors,
as well as spin-base transformations
\cite{Schroedinger:1932a,Bargmann:1932a,Weldon:2000fr,Gies:2013noa} 
\begin{equation}
    G_{\mu\nu}\to \mathcal{S} G_{\mu\nu} \mathcal{S}^{-1}, \quad \mathcal{S}\in 
   \text{SL}(32,\mathbb{C}).
   \label{eq:SBtrafo}
\end{equation}
Analogous to the conventional way of defining Lorentz transformations of, e.g.,
Dirac spinors (leaving the Dirac matrices constant), we can identify the Lorentz
transformations $S_{\text{Lor}}$ of Luttinger spinors as the subgroup of the
spin-base group $\text{SL}(32,\mathbb{C})$ which rotates the Lorentz transformed
$G_{\mu\nu}$ matrices back to their original constant forms. This implies the
identity
\begin{equation}
    S_{\text{Lor}}^{-1}G_{\mu\nu}S_{\text{Lor}}= G_{\kappa\lambda} 
    \Lambda^\kappa{}_\mu \Lambda^\lambda{}_\nu,
\end{equation} 
Correspondingly, $\psi \to S_{\text{Lor}} \psi$ and $ \bar\psi \to \bar\psi
S_{\text{Lor}}^{-1}$ denote the Lorentz transformation of Luttinger spinors.

Let us start now with the standard form of the mass term $\sim \bar{\psi}\psi$
first discussed in \cite{Gies:2023cnd}, leading to a free Lagrangian of the form
\begin{equation}
    \mathcal{L}= -\bar\psi G_{\mu\nu}\partial^\mu \partial^\nu \psi 
    - m^2 \bar\psi \psi.
    \label{eq:standardmass}
\end{equation}
This mass term is invariant under spin-base and thus also under Lorentz
transformations and real as a consequence of the spin metric being hermitean
$h=h^\dagger$. 

The corresponding equation of motion for the field $\psi$ reads in momentum
space
\begin{equation}
    (G_{\mu\nu}p^\mu p^\nu - m^2)\psi = 0.
\end{equation}
Multiplying by $(G_{\kappa\lambda}p^\kappa p^\lambda + m^2)$ from the left
yields
\begin{equation}\begin{split}
    &(G_{\kappa\lambda}p^\kappa p^\lambda + m^2)(G_{\mu\nu}p^\mu p^\nu - m^2)
    \psi  \\
    &\quad = (p^4 - m^4) \psi = (p^2 - m^2)(p^2 + m^2) \psi =0,
\end{split}
\end{equation}
In addition to the expected massive relativistic dispersion relation $p^2=m^2$,
this mass term also gives rise to tachyonic solutions with $p^2=-m^2$. An
explicit check confirms that both types of solutions occur with multiplicity 16
\cite{Schiffhorst:2024}.

A second possible local fermionic bilinear is given by $\bar{\psi}\g_{10} \psi$.
In order to add such a term in a way that the action stays real, it is
instructive to verify the hermiticity properties of this bilinear. We observe
that 
\begin{eqnarray}
    (\bar\psi \g_{10} \psi)^\dagger &=& \psi^\dagger \g_{10}^\dagger 
    \bar\psi^\dagger = \psi^\dagger \g_{10} h \psi = - \psi^\dagger h \g_{10} 
    \psi \nonumber\\
    &=& - \bar\psi \g_{10} \psi,
\end{eqnarray}
where we have used the unitarity of $\g_{10}$ and $h$ as well as the explicit
form of our choice for $h$ in \Eqref{eq:spinmetric}. Therefore, reality of the
action implies to choose a Lagrangian of the form
\begin{equation}
    \mathcal{L}= -\bar\psi G_{\mu\nu}\partial^\mu \partial^\nu \psi 
    - i m^2_{10} \bar\psi \g_{10} \psi. \label{eq:mass10}
\end{equation}
The equation of motion in momentum space reads
\begin{equation}
    (G_{\mu\nu}p^\mu p^\nu - i m^2_{10} \g_{10})\psi = 0.
\end{equation}
Since all $G_{\mu\nu}$ anticommute with $\g_{10}$, we multiply the equation of
motion by $(G_{\kappa\lambda}p^\kappa p^\lambda - i m^2_{10} \g_{10})$ and find 
\begin{equation}
    \begin{split}
    &(G_{\kappa\lambda}p^\kappa p^\lambda  
    - i m^2_{10} \g_{10})(G_{\mu\nu}p^\mu p^\nu  - i m^2_{10} \g_{10})\psi  \\
    &\quad =(p^4 - m^4_{10}) \psi = (p^2 - m^2_{10})(p^2 + m^2_{10}) \psi =0,
    \end{split}
\end{equation}
i.e., we again obtain solutions with both a regular massive as well as a
tachyonic dispersion relation; also the corresponding multiplicities are 16
modes each, as for the standard mass term above. In fact, this is not
astonishing, since both mass terms are connected by a discrete chiral/axial
transformation. For this we first note, that the kinetic term \eqref{eq:Sfree}
features a continuous $\text{U(1)}_{10}$ symmetry,
\begin{equation}
    \psi\to e^{i\vartheta\gamma_{10}}\psi,\quad 
    \bar\psi\to\bar\psi   e^{i\vartheta\gamma_{10}}, \label{eq:axial10}
\end{equation}
which is broken by each of the mass terms discussed above. However, starting
from the massive theory \eqref{eq:standardmass} and performing a
$\text{U(1)}_{10}$ transformation \eqref{eq:axial10} with the choice
$\vartheta=\frac{\pi}{4}$, we obtain the Lagrangian \eqref{eq:mass10} upon the
identification $m^2 \to m_{10}^2$. This also explains, why the solution spectra
and multiplicities match upon this identification. 

The situation is somewhat analogous to conventional Dirac theory, where mass
terms of the form $-m\bar\psi\psi$ and $-m\bar\psi \gamma_5 \psi$ are connected
by an analogous discrete axial transformation. 

Next, we can also use the 11th Euclidean Dirac matrix $\g_{11}$ in order to form
a bilinear mass term. Using the fact that $[h,\g_{11}]=0$, we can verify the
reality property
\begin{equation}
    (\bar\psi \g_{11} \psi)^\dagger = \psi^\dagger \g_{11} h \psi 
    =  \psi^\dagger h \g_{11} \psi = \bar\psi \g_{11} \psi.
\end{equation}
The corresponding free Lagrangian now reads
\begin{equation}
    \mathcal{L}= -\bar\psi G_{\mu\nu}\partial^\mu \partial^\nu \psi 
    - m^2_{11} \bar\psi \g_{11} \psi, \label{eq:mass11}
\end{equation}
giving rise to the equation of motion
\begin{equation}
    (G_{\mu\nu}p^\mu p^\nu - m^2_{11} \g_{11})\psi = 0.
\end{equation}
Since $G_{\mu\nu}$ anticommutes with $\g_{11}$, we multiply by
$(G_{\kappa\lambda}p^\kappa p^\lambda - m^2_{11} \g_{11})$, yielding this time
\begin{equation}
    \begin{split}
    &(G_{\kappa\lambda}p^\kappa p^\lambda  - m^2_{11} \g_{11})
    (G_{\mu\nu}p^\mu p^\nu  - m^2_{11} \g_{11})\psi  \\
    &\quad = (p^4 + m^4_{11}) \psi =0.
    \end{split}
\end{equation}
In contrast to the previous cases, the dispersion relation is now solved by two
complex zeros $p^2=\pm i m_{11}^2$. Both types occur with multiplicity 16.
Neither a standard massive nor a tachyonic mode are present. It is interesting
to note that the Lagrangian \eqref{eq:mass11} is invariant under
$\text{U(1)}_{10}$ transformations \eqref{eq:axial10} as well.

Finally, we can use a product of the matrices $\g_{10}$ and $\g_{11}$ to
construct another independent bilinear, for which we also check its reality
properties based on the identities used above,
\begin{eqnarray}
    (\bar\psi \g_{10}\g_{11} \psi)^\dagger 
    &=& \psi^\dagger \g_{11}\g_{10} h \psi  
    = - \psi^\dagger \g_{10}\g_{11} h \psi \nonumber\\  
    &=&  \psi^\dagger h \g_{10}\g_{11} \psi 
    = \bar\psi \g_{10}\g_{11} \psi.
\end{eqnarray}
For convenience, let us introduce the hermitean product
\begin{equation}
    \g_{01}:= -i \g_{10} \g_{11}, \quad \g_{01}^\dagger=\g_{01},
    \label{eq:gamma01}
\end{equation}
which satisfies
\begin{equation}
    [G_{\mu\nu},\g_{01}]=0,\,\, \{h,\g_{01}\}=
    \{\g_{10},\g_{01}\}=\{\g_{11},\g_{01}\}=0.
\end{equation}
Correspondingly, the free real Lagrangian containing the new bilinear can be
written as
\begin{equation}
    \mathcal{L}= -\bar\psi G_{\mu\nu}\partial^\mu \partial^\nu \psi 
    - i m^2_{01} \bar\psi \g_{01} \psi, \label{eq:mass01}
\end{equation}
yielding the equation of motion
\begin{equation}
    (G_{\mu\nu}p^\mu p^\nu - i m^2_{01} \g_{01})\psi = 0.
\end{equation}
As $G_{\mu\nu}$ commutes with $\g_{01}$, we multiply by
$(G_{\kappa\lambda}p^\kappa p^\lambda + i m^2_{01} \g_{01})$ and obtain
\begin{equation}
    \begin{split}
    &(G_{\kappa\lambda}p^\kappa p^\lambda  + i m^2_{01} \g_{01})
    (G_{\mu\nu}p^\mu p^\nu  - i m^2_{01} \g_{01})\psi  \\
    &\quad =(p^4 + m^4_{01} \g_{01}^2)\psi  
     =(p^4 + m^4_{01}) \psi =0,
    \end{split}
    \label{eq:m01poles}
\end{equation}
since $\g_{01}$ squares to one. As in the preceding case, we observe complex
conjugate zeros in the momentum plane $p^2$, implying solutions with a
dispersion relation $p^2=\pm i m_{01}^2$. Each type of solution has again
multiplicity 16. Also, the Lagrangian \eqref{eq:mass01} is invariant under the
$\text{U(1)}_{10}$ symmetry.

It is tempting to expect that each of the dispersion relations found for
the different free massive theories corresponds to a generic analytic pole
structure in the complex $p^2$ plane. Whether or not this is the case is
discussed in Sect.~\ref{sec:propagators}.

It is suggestive to introduce two further U(1) transformations, namely,
\begin{eqnarray}
    \text{U(1)}_{11}: &\,\,& \psi \to e^{i\vartheta \g_{11}} \psi, \,\,
        \bar\psi \to  \bar\psi e^{-i\vartheta \g_{11}},\label{eq:U11} \\
    \text{U(1)}_{01}: &\,\,& \psi \to e^{i\vartheta \g_{01}} \psi, \,\,
        \bar\psi \to  \bar\psi e^{i\vartheta \g_{01}}.\label{eq:U01}
\end{eqnarray}
We observe that the four mass terms can be connected via discrete versions of
these transformations: e.g., the $m_{01}$ mass is connected to the $m_{10}$ mass
via a $\text{U(1)}_{11}$ transformation with $\vartheta=\frac{\pi}{4}$.

However, it is important to emphasize that the transformations \eqref{eq:U11}
and \eqref{eq:U01} do not represent symmetries of the kinetic term and thus are
no symmetries of the Lagrangians if taken at face value. Some of these
transformations may, nevertheless, be uplifted to a symmetry, if combined with a
simultaneous transformation of the spin metric. E.g., we observe that a discrete
$\text{U(1)}_{01}$ transformation with $\vartheta=\frac{\pi}{4}$ transforms the
kinetic term into an analogous kinetic term with the spin metric being replaced
by $h\to \g_{1} \g_2 \g_3 \g_{11}$. The latter is also a valid choice for the
spin metric satisfying all necessary conditions of \Eqref{eq:hcond}.

The existence of a set of different mass-like terms is similar to that for Dirac
fermions in reducible representation, with the $d=3$ case with $d_\gamma=4$
being the most well-studied case
\cite{Semenoff:1984dq,Appelquist:1986fd,Haldane:1988zza,Kane:2004bvs}. In
contrast to this, the present case of Luttinger fermions is not a reducible
representation: though the Abrikosov algebra \eqref{eq:AbrikosovA} could be
represented by 16-dimensional matrices in $d=4$, the spin metric cannot and thus
requires a 32-dimensional representation. From a technical viewpoint the
properties of the spin metric are also responsible for the fact that the
transformations \eqref{eq:U11} and \eqref{eq:U01} do not correspond to
symmetries of the action. Hence, there is also no extended flavor symmetry such
as U(2$\Nf$) as in the case of $d=3$ reducible Dirac fermions.

Let us finally remark that the existence of further mass terms is
conceivable; e.g, if the fermionic field satisfies additional reality
constraints, mass terms analogous to Majorana masses in the
Dirac case may be allowed.

\section{Self-interacting fermionic models}\label{sec:selfintmodels}

Let us introduce a set of massless theories of self-interacting relativistic
Luttinger fermions with interactions defined in terms of the above-mentioned
spinor bilinears. For a first glance at the quantum theory, we perform a
one-loop analysis of their RG flow concentrating on the large-$\Nf$ limit for
simplicity. 

In the present section, we work in the Euclidean domain in order to use
Wilsonian RG techniques. Note that the definition of the following models in the
Euclidean differs by a minus sign in the interaction terms from the formulation in 
Minkowskian spacetime, cf. App.~\ref{sec:AppC}. 

We start with the simplest interaction term which is reminiscent to that of the
standard Gross-Neveu \cite{Gross:1974jv} model, cf. \cite{Gies:2023cnd}
\begin{equation}\label{eqn:SGNE}
S=\int d^4x \left[- \bar\psi G_{\mu\nu}\partial^\mu \partial^\nu \psi + 
\frac{\bar\lambda}{2} (\bar\psi \psi)^2\right].
\end{equation}
Flavor indices are suppressed for simplicity here and in the following; all
bilinears written in terms of parentheses are assumed to be flavor singlets,
i.e., $(\bar\psi \psi)= \bar\psi^a \psi^a$, where $a=1,\dots,N_f$. This, as well as all subsequent
models, therefore features a global U($\Nf$) symmetry which will remain trivially
present in all of the subsequent discussion. 

The Luttinger-Gross-Neveu model additionally exhibits a discrete axial symmetry
of the type of \Eqref{eq:axial10} with the choice $\vartheta= \frac{\pi}{2}$.
Under such a transformation, the kinetic term is invariant. The scalar bilinear
transforms as $(\bar\psi\psi) \to -(\bar\psi \psi)$ which leaves the interaction
term in \Eqref{eqn:SGNE} invariant, but forbids the occurrence of a mass term.
In analogy to the standard Gross-Neveu model, it is tempting to speculate that
this discrete symmetry might be broken depending on the sign and the strength of
the initial value for the coupling $\bar{\lambda}$. 

Another rather similar model is given by a scalar interaction involving the  $\g_{10}$ matrix,
\begin{equation}\label{eqn:S10E}
S=\int d^4x \left[- \bar\psi G_{\mu\nu}\partial^\mu \partial^\nu \psi - 
\frac{\bar\lambda}{2} (\bar\psi \g_{10}\psi)^2\right].
\end{equation}
Here and in the following, the sign in front of the coupling is chosen such that
the one-loop beta functions computed below have the same form. Also, we use the
same letter $\bar\lambda$ for the coupling for simplicity, even though the
couplings in all the models considered here are unrelated. Also this
$\g_{10}$ model has the same discrete axial symmetry as the
Luttinger-Gross-Neveu model: under the transformation \eqref{eq:axial10} with
the choice $\vartheta= \frac{\pi}{2}$, the Lagrangian in \Eqref{eqn:S10E}
remains invariant, whereas a mass term of the $m_{10}$ type as in
\Eqref{eq:mass10} would change sign and thus break the symmetry. 

Next, we introduce the $\g_{11}$ model in terms of the action
\begin{equation}\label{eqn:S11E}
S=\int d^4x \left[- \bar\psi G_{\mu\nu}\partial^\mu \partial^\nu \psi - 
\frac{\bar\lambda}{2} (\bar\psi \g_{11}\psi)^2\right],
\end{equation}
This $\g_{11}$ model is invariant under the full continuous $\U(1)_{10}$
symmetry \eqref{eq:axial10}. However, already the $m_{11}$ mass term in
\Eqref{eq:mass11} is invariant under this symmetry, hence the realization of
this symmetry does not serve as an indicator for gap formation. Instead, this
role is played by a combined discrete symmetry involving both a discrete
$\U(1)_{11}$ transformation \eqref{eq:U11} with $\vartheta=\frac{\pi}{2}$ and
the replacement $\psi\to-\psi$ and $\bar\psi \to \bar\psi$ (treating $\psi$ and
$\bar\psi$ as independent variables in the quantum theory). The $\g_{11}$ model
\eqref{eqn:S11E} is invariant under this discrete transformation while an
$m_{11}$ mass term is not. This discrete symmetry is somewhat similar to the
discrete symmetry of the 3d Gross-Neveu model with irreducible Dirac fermions
\cite{Hofling:2002hj}. 

As a fourth action, we consider the $\g_{01}$ model:
\begin{equation}\label{eqn:S01E}
S=\int d^4x \left[- \bar\psi G_{\mu\nu}\partial^\mu \partial^\nu \psi + 
\frac{\bar\lambda}{2} (\bar\psi \g_{01}\psi)^2\right].
\end{equation}
Also the $\g_{01}$ model is invariant under  the full continuous $\U(1)_{10}$
symmetry \eqref{eq:axial10}, as we observed already for the $m_{01}$ mass term
in \Eqref{eq:mass01}. Hence, the status of this symmetry is not indicative for
mass generation. In fact, none of the transformations discussed in the
previous sections is a suitable ingredient for constructing an indicator
symmetry for mass gap formation as each of them acts similarly on the kinetic
and the mass term. Still, we have checked explicitly that the interaction does
not generate an $m_{01}$ mass term at one-loop order. This implies that either a
mass-protecting symmetry exists or an $m_{01}$ mass term may be generated at
higher-loop order.

Finally, we note that the models can, of course, also be combined such that
continuous symmetries emerge. An example is given by a Luttinger-fermionic
analogue of the Nambu$-$Jona-Lasinio (NJL) model \cite{Nambu:1961tp}, which
features a full continuous  $\U(1)_{10}$ symmetry \eqref{eq:axial10}, as first
discussed in \cite{Gies:2023cnd},
\begin{equation}\label{eqn:SNJLE}
S=\int d^4x \left\{- \bar\psi G_{\mu\nu}\partial^\mu \partial^\nu \psi + 
\frac{\bar\lambda}{2} \left[(\bar\psi \psi)^2-(\bar\psi \g_{10}\psi)^2\right]\right\}.
\end{equation}
In this model, the $\U(1)_{10}$ symmetry forbids corresponding mass terms such that the status of the symmetry can be expected to be indicative of gap formation. 

For each of these theories, we compute the one-loop beta function. While this
can straightforwardly be done with any conventional quantum field theory method,
we use the functional renormalization group (RG) here, as it can be generalized
straightforwardly to future nonperturbative studies. Specifically for fermionic
theories, the computational techniques based on the Wetterich equation
\cite{Wetterich:1992yh} are well developed
\cite{Gies:2001nw,Braun:2011pp,Gehring:2015vja} and have found manifold
nonperturbative applications
\cite{Hofling:2002hj,Braun:2010tt,Mesterhazy:2012ei,Jakovac:2014lqa,
Janssen:2014gea,Vacca:2015nta,Classen:2015mar,Knorr:2016sfs,Cresswell-Hogg:2022lgg,
Cresswell-Hogg:2022lez,Cresswell-Hogg:2023hdg}. Starting from the Wetterich
equation for the effective average action  $\Gamma_k$,
\begin{equation}
 \pat \Gamma_k= \frac{1}{2} \STr \big[\pat R_k 
(\Gamma_k^{(2)}+R_k)^{-1}\big],
 \label{eq:Wetterich}
\end{equation}
the regulator function $R_k$ implements the regularization in the Euclidean momentum domain at a regularization scale $k$; here $\pat=k \frac{d}{dk}$. Provided $\Gamma_k$ is fixed in terms of the bare microscopic action $S$ for $k\to\Lambda$ as a UV boundary condition, the full quantum effective action  $\Gamma$ is approached for $k\to 0$ in the IR. Importantly, it can be chosen in such a way that the symmetries of the kinetic term are respected by the regularization procedure. Using this as well as standard methods as detailed in \cite{Gies:2023cnd}, we project the Wetterich equation for each of the models onto a theory space defined by the ansatz
\begin{equation}
\Gamma_k=\int_x \left[-Z_\psi \bar\psi G_{\mu\nu}\partial^\mu \partial^\nu \psi + \mathcal{L}_{\text{int}}\right],
\label{eq:truncGammak}
\end{equation}
where $\mathcal{L}_{\text{int}}$ denotes the interaction term of the corresponding model including the scale-dependent coupling $\bar\lambda$ and a wave function renormalization $Z_\psi$. Introducing the renormalized coupling 
\begin{equation}
\lambda = \frac{\bar\lambda}{Z_\psi^2}, \label{eq:lambda}
\end{equation}
we find for each of the five models the beta function
\begin{equation}\label{eqn:betafunctionFRG}
\pat \lambda = - \frac{4 N_f}{\pi^2} \lambda^2
\end{equation}
in the large-$N_f$ limit. (To one-loop order, the anomalous dimension $\eta_\psi = - \pat \ln Z_\psi$ vanishes $\eta_\psi=0$, which completes the flow in the theory space spanned by the ansatz \eqref{eq:truncGammak}.) Equation~\eqref{eqn:betafunctionFRG}  demonstrates that each of these models is asymptotically free for positive $\lambda>0$, approaching the Gaussian fixed point towards the ultraviolet (UV) as a high-energy fixed point. Asymptotic freedom guarantees that the models can be extended to arbitrarily high energy scales. Towards low energies, the coupling $\lambda$ grows larger and the true behavior of the models has to be analyzed by nonperturbative means.  

By contrast, the Gaussian fixed point is infrared attractive for negative couplings, $\lambda<0$. Correspondingly, the couplings diverge to negative infinity towards high energies (Landau poles); thus, a nonperturbative analysis is necessary to search for a possible UV completion or to prove triviality of the models in this coupling branch.  

Of course, the present analysis can straightforwardly be generalized to finite $\Nf$ values. However, a consistent treatment in this regime requires to include a Fierz-complete set of interaction channels. This has, e.g., been done for the Luttinger-Gross-Neveu model in \cite{Heinzel:2023,Gies:2023cnd} which required the inclusion of a tensor channel $\sim (\bar\psi G_{\mu\nu} \psi)^2$. The property of coupling branches where the theory is asymptotically safe then generalizes to higher dimensional regions in the space of all couplings. We expect similar properties to hold for each of the models studied here.

\section{Mean-field theory}
\label{sec:MFT}

In order to investigate the possible occurrence of gap formation in the models
defined in the previous section, we use mean-field theory which becomes exact in
the large-$\Nf$ limit. For this, we bilinearize the fermionic actions given
above using auxiliary scalar fields with a Gaussian action and a Yukawa coupling
to the fermionic fields. In the large-$\Nf$ limit, the auxiliary scalar field
integral is dominated by the classical configurations, i.e., the extrema of the
action which in turn is governed by the fermion determinant. Since the
true expansion parameter of the large-$\Nf$ limit also involves the
dimensionality of the Clifford algebra \cite{Vacca:2015nta}, the expansion is in
powers of $\frac{1}{d_\gamma \Nf}= \frac{1}{32\Nf}$ which is already a small
parameter for $\Nf=1$.

In view of various forms of possible mass terms discussed in
Sect.~\ref{sec:massterms}, we expect radiatively generated gaps to occur in the
complex momentum plane. Therefore, we perform the mean-field analysis in
Minkowski space, using propertime methods for the regularization. Of course, for
all models, the sign change of the interaction term when comparing the Euclidean
description used in Sect.~\ref{sec:selfintmodels} with the present Minkowskian
analysis has to be accounted for, cf. App.~\ref{sec:AppC}. 

\subsection{Luttinger-Gross-Neveu model}

Let us start with the Luttinger-Gross-Neveu model, the action of which in
Minkowski spacetime reads
\begin{equation}\label{eqn:SGNM}
S=\int d^4x \left[-\bar\psi G_{\mu\nu}\partial^\mu \partial^\nu \psi 
- \frac{\bar\lambda}{2} (\bar\psi \psi)^2\right].
\end{equation}
We bilinearize the interaction term using a Hubbard-Stratonovich (HS)
transformation introducing an auxiliary real scalar field, such that the action
reads
\begin{equation}
S_{\text{FB}}=\int d^4x \left[-Z_\psi \bar\psi G_{\mu\nu}\partial^\mu \partial^\nu \psi 
+ \barh \phi \bar\psi \psi -\frac12 \bar{m}^2 \phi^2\right],
\label{eqn:SFBGN}
\end{equation}
First, we note that the action \eqref{eqn:SFBGN} is manifestly real if the
coupling $\barh$ is real, since both $\phi$ and $\bar\psi \psi$ are real. Also,
the sign of the scalar mass term is such that it corresponds to a positive mass
term and thus a stable potential in Minkowski space. The discrete axial symmetry
of the fermionic description is also preserved by the action \eqref{eqn:SFBGN}
if the scalar field transforms as $\phi\to - \phi$. The theories defined by
\Eqref{eqn:SGNM} and \Eqref{eqn:SFBGN} are identical both on the classical as
well as on the quantum level, provided the coupling constants satisfy a matching
condition. The matching condition can, e.g., be derived from the classical
equation of motion for the scalar field (corresponding to a Gaussian integration
on the quantum level) which reads
\begin{equation}
\frac{\delta S}{\delta \phi} = \barh \bar\psi \psi - \bar{m}^2 \phi = 0,
\label{eq:HSclassical}
\end{equation}
Inserting the solution for $\phi$ into \Eqref{eqn:SFBGN} leads us back to
\Eqref{eqn:SGNM} provided the matching condition
\begin{equation}\label{eqn:HSrelation}
\bar\lambda = - \frac{\barh^2}{\bar{m}^2}
\end{equation}
is satisfied. Incidentally, the Yukawa coupling could be set to a unit scale  by rescaling the scalar field; we keep it for reasons of generality. The minus
sign in \eqref{eqn:HSrelation} implies that the HS transformation can be
meaningfully performed in the standard fashion only for negative values of the
coupling $\bar\lambda$. This confines the following analysis to the
non-asymptotically free branch $\bar\lambda<0$ of the Luttinger-Gross-Neveu
model. While this corresponds to the branch where the Gaussian fixed point is IR
attractive, we are still free to assume that the initial value of the coupling
$\bar\lambda$ is sufficiently large to potentially introduce a nontrivial
long-range behavior.

For this, we investigate the effective action of the scalar field upon
integrating out the fermions. This functional integral yields the one-loop
contribution $\Gamma_{1\ell}$ to the effective action in terms of the fermion
determinant. We evaluate the latter for a constant scalar field,
$\phi=\phi_0=$constant:
\begin{eqnarray}
\Gamma_{1\ell} &=& -i \ln\det [-G_{\mu\nu} \partial^\mu \partial^\nu 
+ \barh \phi_0] \nonumber\\
&=& -\frac{i}{2} \ln\det [- (\partial^2)^2 + (\barh \phi_0)^2],
\label{eq:Gamma1ellLGN}
\end{eqnarray}
where in the last step we have used the $\gamma_{10}$ hermiticity of the kinetic
term, $\gamma_{10} G_{\mu\nu}\partial^\mu \partial^\nu \gamma_{10}=
-G_{\mu\nu}\partial^\mu \partial^\nu$. Now, we employ $\ln \det = \Tr \ln$ and perform a trivial vacuum subtraction such that $\Gamma_{1\ell}[\phi_0=0]=0$. Going to Fourier space and evaluating the functional trace we arrive at 
\begin{eqnarray}
\Gamma_{1\ell} &=& -\frac{i}{2} N_f d_{\g} \Omega \int \frac{d^4p}{(2\pi)^4} 
\ln\left(\frac{p^4-(\barh\phi_0)^2}{p^4}\right) \nonumber\\
&=& \frac{N_f d_{\g} \Omega}{2^6 \pi^2} \int \frac{dt}{t^2} 
\left( 1-e^{(\barh\phi_0)^2t}\right).
\label{eq:Gam1ell}
\end{eqnarray}
For the last step, we have rotated the momentum integral to the Euclidean and
used the Schwinger propertime representation of the logarithm. This
representation is both infrared and ultraviolet divergent. We can cure the UV
divergence with the introduction of a UV cut-off scale $\Lambda$, i.e.,
introduce a lower bound of the integral at $1/\Lambda^4$. Of course, the UV
divergence is indicative for the renormalization of the couplings. 

The IR divergence signals the existence of tachyonic modes. This is already
obvious from the first line of \Eqref{eq:Gam1ell} where the argument of the
logarithm becomes negative for momenta with $p^4< (\barh \phi_0)^2$. We deal
with the IR divergence by studying the integral in the complex $(\barh\phi_0)^2$
plane where it exists for all $\text{Re}(\barh\phi_0)^2<0$; we then continue the
result analytically back to real values of $\phi_0$. As a result, the effective
action picks up an imaginary part indicating that the assumption of a finite
scalar mean field $|\phi_0|>0$ would correspond to an unstable vacuum state.
Expanding the resulting expression in inverse powers of the UV cutoff $\Lambda$,
we obtain
\begin{eqnarray}
\Gamma_{1\ell} &=& -\frac{N_f d_{\g} \Omega}{2^6 \pi^2} (\barh\phi_0)^2 
\left[1-\g-\ln\left(\frac{(\barh\phi_0)^2}{\Lambda^4}\right)-i\pi\right]\nonumber\\
&& + O((\barh \phi_0)^2/\Lambda^4),
\label{eqn:G1lGN}
\end{eqnarray}
where $\Omega$ denotes the spacetime volume. Following Schwinger
\cite{Schwinger:1951nm}, the imaginary part of the effective action is a measure
for the decay rate of a state with finite $\phi_0$ with $\exp(-2\text{Im}
\Gamma)$ quantifying the probability for the state to persist. This tells us
already that within our assumptions only the $\phi_0=0$ state can be an
equilibrium state. 

In order to further study the stability of this state, we consider the (real
part of the) effective mean-field potential including the classical scalar mass
term but ignoring all terms that vanish in the limit $\Lambda\to\infty$,
\begin{equation}\label{eqn:veffGN}
V_{\text{eff}}(\phi_0)=\frac{1}{2}\phi_0^2 \left\{\bar{m}^2 +
 \frac{N_f d_{\g}}{2^5 \pi^2}\barh^2 
 \left[1-\g-\ln\left(\frac{(\barh\phi_0)^2}{\Lambda^4}\right)\right] \right\},
\end{equation}
where $\gamma\simeq 0.5772\dots$ denotes the Euler-Mascheroni constant.
With $\Lambda$ being the largest scale (to be sent to infinity), we observe
already in this unrenormalized expression that the term dominating 
the effective potential at large fields is positive, $V_{\text{eff}}(\phi_0) \sim - \phi_0^2 \ln
[(\barh \phi_0)^2 /\Lambda^4] > 0$. Also, all other terms are positive for finite
$\phi_0$ and vanish only for $\phi_0=0$ in the validity regime of
\Eqref{eqn:veffGN} with $\Lambda^2\gg (\barh \phi_0)$. Therefore, the zero-field
mean-field state $\phi_0=0$ is the minimum of the (real part of the) effective
potential. 

\begin{figure}[t]
\includegraphics[width=0.48\textwidth]{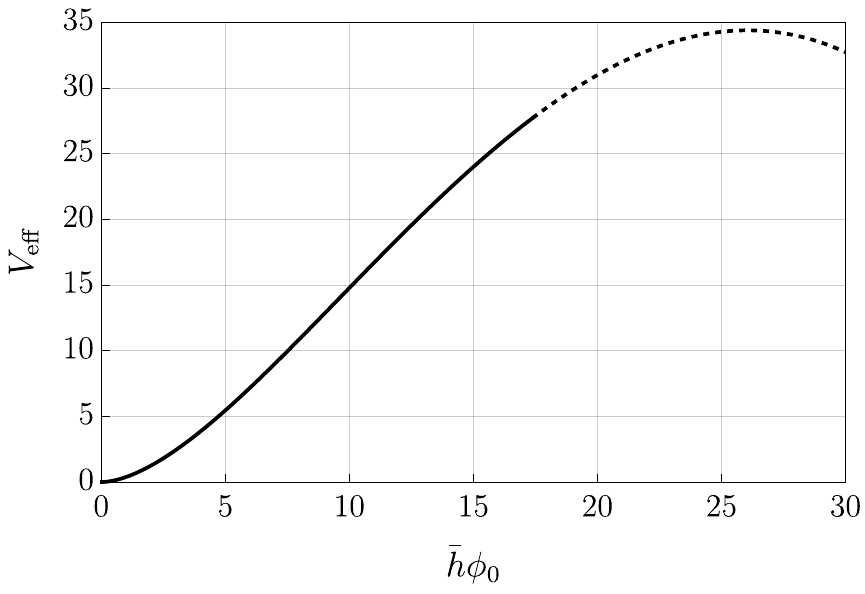}
\caption{Renormalized effective potential $V_{\text{eff}}$ of the
Luttinger-Gross-Neveu model for $\Nf=1$ and $d_{\gamma}=32$. The plot is
obtained by setting the renormalization scale to a small value, namely $\mu=0.5$
and the ratio $m^2(\mu)/\barh^2$ is set to $1$. Moreover, since the mean-field
analysis is done for (possibly large) negative values of the bare
coupling $\bar{\lambda}$, we set the bare mass parameter to its lower bound,
i.e. $\bar{m}=0$ (larger values would correspond to couplings closer to
zero). Choosing $\barh\phi_0<0.5 \Lambda^2$ as an ad hoc criterion for the
validity regime of our analysis requiring, in principle, $|\barh\phi_0|
\ll \Lambda^2$, the effective potential is depicted with a solid line in the
validity region, and with a dashed line where the assumptions are
violated.}
\label{fig:XY}
\end{figure}

Of course, we can also introduce renormalized quantities by defining a
renormalized mass at some renormalization scale $\mu$, 
\begin{equation}
    m^2(\mu)\coloneqq \bar{m}^2 -\frac{N_f d_{\g}}{2^5 \pi^2}\barh^2 
    \ln\frac{\mu^4}{\Lambda^4}.
    \label{eq:GNrenmass}
\end{equation}
We emphasize that both terms on the right-hand side are strictly positive, since
$\mu \ll \Lambda$ for a meaningful renormalization scale well below the UV
cutoff. The correspondingly renormalized effective potential reads
\begin{equation} \label{eqn:veffGNmu}
V_{\text{eff}}(\phi_0)=\frac{\phi_0^2 }{2}\left\{m^2(\mu) 
+ \frac{N_f d_{\g}}{2^5 \pi^2}\barh^2 \left[1-\g
-\ln
\frac{(\barh\phi_0)^2}{\mu^4}
\right] \right\},
\end{equation}
For small fields $(\barh\phi_0)^2<\mu^4$, we again observe that the term in
curly brackets remains positive, hence $\phi_0=0$ is a local minimum of the
effective potential. On the other hand, for large fields
$(\barh\phi_0)^2\gg\mu^4$, it naively seems that the effective potential is not
bounded from below, since the logarithm can become arbitrarily large. However,
this is an artifact of this representation. As discussed above, the mean-field
potential in the representation \eqref{eqn:veffGN} stays positive for all
$\bar{m}^2>0$, since $\Lambda$ is the largest scale in the game and eventually
goes to infinity. Indeed, for $\bar{m}^2>0$, $m^2(\mu)$ grows logarithmically
for decreasing $\mu$, which compensates the logarithmic increase of
$\ln\left(\frac{(\barh\phi_0)^2}{\mu^4}\right)$, keeping \eqref{eqn:veffGNmu}
positive. A plot of the effective potential $V_{\text{eff}}$ is depicted
in Fig.~\ref{fig:XY}, confirming that the trivial vacuum $\phi_0=0$ is the
global minimum in the validity range of the computation (solid line).

The present discussion is very
similar to that of mean-field or Coleman-Weinberg-type effective potentials in
Dirac-Yukawa theories \cite{Holland:2004sd,Gies:2013fua,Gies:2014xha}, where a
naive look at the renormalized form can be misleading if the ultimate existence
of a UV cutoff is ignored. 

In summary, we conclude that $\phi_0=0$ is not only a local, but the global
minimum of $V_{\text{eff}}$ in the mean-field approximation for the
$\bar\lambda<0$ branch of the Luttinger-Gross-Neveu model. At mean field level,
the fermions develop neither a mass nor a tachyonic mode in this branch of the
model even if the initial value of the bare coupling $\bar\lambda<0$ has a large
absolute value at the UV scale $\Lambda$. Since the standard HS transformation
cannot be applied to the positive coupling branch, we obtain no information
about the status of the model in the asymptotically free branch where the couplings grows towards the IR.

As a consistency check, we can compute the $\beta$ function at mean-field level
by introducing the scale-dependendent coupling $\lambda(\mu)=-\barh^2/m^2(\mu)$,
which for $d_\gamma=32$ results in
\begin{equation}\label{eqn:betafunctionGN}
\mu \frac{\partial}{\partial\mu} \lambda(\mu) = - \frac{4N_f}{\pi^2}\lambda^2(\mu).
\end{equation}
Strictly speaking, we have derived this result for negative values of $\lambda$
only. But it agrees with the result \eqref{eqn:betafunctionGN} of the preceding
section for all values of $\lambda$, thereby reproducing the one-loop RG flow in
the large-$\Nf$ limit including asymptotic freedom in the positive coupling
branch. 

\subsection{$\g_{10}$ model}

Let us now study the second model discussed above, with Minkowskian
action
\begin{equation}\label{eqn:S10M}
S=\int d^4x \left[- \bar\psi G_{\mu\nu}\partial^\mu \partial^\nu \psi 
+ \frac{\bar\lambda}{2} (\bar\psi \g_{10}\psi)^2\right].
\end{equation}
As before, we aim at the mean-field potential in order to explore the
possibility of gap formation. This time, the HS transformation leads us to
\begin{equation}\label{eqn:SFBg10}
S_{\text{FB}}=\int d^4x \left[- \bar\psi G_{\mu\nu}\partial^\mu \partial^\nu \psi 
+ i \barh \phi \bar\psi \g_{10} \psi -\frac12 \bar{m}^2 \phi^2\right],
\end{equation}
for the partially bosonized version of the model. The factor of $i$ in front of
the Yukawa interaction guarantees that the action is real, cf.
\Eqref{eq:mass10}. The discrete axial symmetry of \Eqref{eqn:S10M} again induces
a $\mathbb{Z}_2$ symmetry for the scalar $\phi\to -\phi$. 

It turns out that the equivalence of the two actions requires the same matching
condition \eqref{eqn:HSrelation}. Again, only the negative coupling branch
$\bar\lambda<0$ can be studied in mean-field theory. 

It is also straightforward to verify that the mean-field analysis leads to the
same one-loop effective action including the imaginary part for finite $\phi_0$
as well as the same effective potential as in the Luttinger-Gross-Neveu model
\eqref{eqn:veffGN}, with a global minimum at $\phi_0=0$. Also in this case, the
Luttinger fermions remain ungapped and do not exhibit a tachyonic mode in the
negative coupling branch. While we again cannot address the long-range physics
in the positive coupling branch, the mean-field analysis yields the correct
$\beta$ function \eqref{eqn:betafunctionGN} for all values of the coupling.

With hindsight, the fact that the two models behave identically is not too
surprising, since the discrete $\U(1)_{10}$ transformation with $\vartheta =
\frac{\pi}{4}$ discussed below \Eqref{eq:axial10} transforms the
Luttinger-Gross-Neveu model into the $\g_{10}$ model at each stage of the
analysis.

\subsection{$\g_{11}$ model}

Now, the $\g_{11}$ model turns out to behave rather differently. We start with
the corresponding action in Minkowski spacetime
\begin{equation}\label{eqn:S11M}
S=\int d^4x \left[- \bar\psi G_{\mu\nu}\partial^\mu \partial^\nu \psi + 
\frac{\bar\lambda}{2} (\bar\psi \g_{11} \psi)^2\right].
\end{equation}
This time, the HS transformation leads us to the partially bosonized action
\begin{equation}
S_{\text{FB}}=\int d^4x \left[- \bar\psi G_{\mu\nu}\partial^\mu \partial^\nu \psi 
+ \barh \phi \bar\psi \g_{11} \psi -\frac12 \bar{m}^2 \phi^2\right]
\label{eqn:SFB11}
\end{equation}
which is equivalent to \eqref{eqn:S11M}, provided the matching condition
\begin{equation}
\bar\lambda = \frac{\barh^2}{\bar{m}^2} \label{eq:matching11}
\end{equation}
is satisfied. We observe that the HS transformation is now tied to the positive
asymptotically free $\bar\lambda$ branch. Again, the discrete symmetry of the
fermionic formulation inhibiting a bare mass term induces a $\mathbb{Z}_2$
symmetry for the scalar field such that $S_{\text{FB}}$ is invariant under the
combined transformation. Choosing the scalar field to be constant
$\phi=\phi_0=$const., the fermion determinant yielding the one-loop contribution
to the effective action reads
\begin{eqnarray}
\Gamma_{1\ell} &=& -i \ln\det [-G_{\mu\nu} \partial^\mu \partial^\nu 
+ \barh \phi_0 \g_{11}] \nonumber\\
&=& -\frac{i}{2} \ln\det [- (\partial^2)^2 - (\barh \phi_0)^2].
\end{eqnarray}
In the last step, we have used the $\gamma_{10}$-hermiticity of the kinetic
term, as well as the anticommutator properties
$\{G_{\mu\nu},\g_{11}\}=\{\g_{10},\g_{11}\}=0$. Evaluating the resulting trace
in Fourier space, performing the vacuum subtraction, and using the propertime
representation, we arrive at
\begin{eqnarray}
    \Gamma_{1\ell} &=& -\frac{i}{2} N_f d_{\g} \Omega \int 
    \frac{d^4p}{(2\pi)^4} 
    \ln\left(\frac{p^4+(\barh\phi_0)^2}{p^4}\right) \nonumber\\
    &=& \frac{N_f d_{\g} \Omega}{2^6 \pi^2} \int \frac{dt}{t^2} 
    \left( 1-e^{-(\barh\phi_0)^2t}\right).
    \label{eq:Gam1ell11}
\end{eqnarray}
In the first line, it is already obvious that a finite value of $\phi_0$ does
neither induce tachyonic modes nor an imaginary part of the action.
Consequently, the propertime representation (second line) requires only a UV
cutoff, implemented by a $1/\Lambda^4$ lower bound at the $t$ integral, whereas
the action is IR finite. Correspondly, the unrenormalized effective potential
including the classical scalar mass term can straightforwardly be computed.
Ignoring the terms that vanish in the large-$\Lambda$ limit, we find
\begin{equation}
V_{\text{eff}}(\phi_0)=\phi_0^2 \left\{\frac{\bar{m}^2}{2} 
- \frac{N_f d_{\g}}{2^6 \pi^2}\barh^2 
\left[1-\g-\ln\left(\frac{(\barh\phi_0)^2}{\Lambda^4}\right)\right] \right\}.
\end{equation}
In order to renormalize the effective potential, we first introduce a
renormalized mass parameter at some renormalization scale $\mu$,
\begin{equation}
    m^2(\mu)\coloneqq \bar{m}^2 +\frac{N_f d_{\g}}{2^5 \pi^2}\barh^2 
    \ln\frac{\mu^4}{\Lambda^4}.
    \label{eq:renmass11}
\end{equation}
Note that $m^2(\mu)$ can take values with either sign in contrast to the
renormalized mass in the previously discussed models, cf. \Eqref{eq:GNrenmass}.
The effective potential can then be written as
\begin{equation}
    V_{\text{eff}}=\frac12 \phi_0^2 \left\{{m}^2(\mu) 
    - \frac{N_f d_{\g}}{2^5 \pi^2}\barh^2 \left[1-\g
    -\ln\left(\frac{(\bar h\phi_0)^2}{\mu^4}\right)\right] \right\}.
\end{equation}
The latter displays a nontrivial minimum satisfying
$V_{\text{eff}}'(\phi_0=v)=0$ at
\begin{equation}
(\barh v)^2 = {\mu^4}\, 
e^{-\frac{2^5\pi^2}{N_f d_{\g}}\frac{m^2(\mu)}{\barh^2}-\gamma}.
\end{equation}
Using \Eqref{eq:renmass11}, it is straightforward to verify that this minimum is
RG invariant,
\begin{equation}
\mu \frac{d}{d\mu} v^2 = 0.
\label{eq:RGinv-v}
\end{equation}
In terms of the minimum, the renormalized potential can also be brought into an
RG invariant form
\begin{equation}
V_{\text{eff}}= \frac{\Nf d_\gamma}{2^6 \pi^2} (\barh \phi_0)^2 
\ln \frac{(\barh\phi_0)^2}{e (\barh v)^2}, 
\label{eq:Veff11v}
\end{equation}
where $e$ denotes the Euler number.

\begin{figure}[t]
\includegraphics[width=0.48\textwidth]{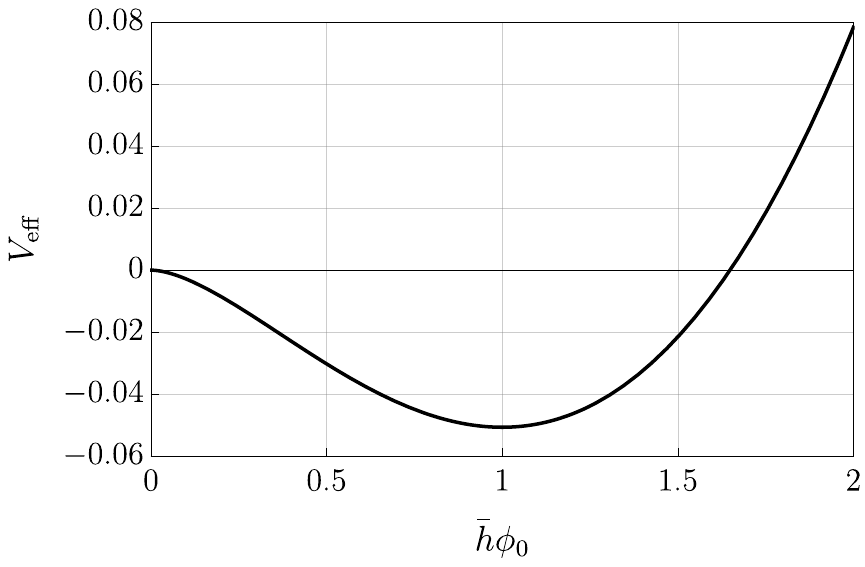}
\caption{Renormalized effective potential of the $\g_{11}$ model for $\Nf=1$ and
$d_{\gamma}=32$. The position of the minimum has been set to $1$, namely $\barh
v=1$, in units of the square of an arbitrary dimensionful scale. The
presence of a minimum at $\phi_0=v$ indicates the formation of a mass gap in the
fermionic spectrum.}
\label{fig:YZ}
\end{figure}

The effective potential is plotted in Fig.~\ref{fig:YZ}; it is bounded from
below and exhibits the nontrivial minimum at $\phi_0=v$. In this
quantum-induced ground state, the discrete $\mathbb{Z}_2$ symmetry is
spontaneously broken, giving rise to a fermionic $m_{11}$ mass term. We can
read off from \Eqref{eqn:SFB11} that
\begin{equation}
m_{11}^2=\barh v. 
\label{eq:m11induced}
\end{equation}
In fact, the product $\barh v$ sets the scale for all dimensionful quantities
occuring in Eqs.~(\ref{eq:Veff11v},~\ref{eq:m11induced}). Since the original theory,
the $\g_{11}$ model, has no intrinsic scale on the classical level, this is a
textbook example for dimensional transmutation. As discussed in
Sect.~\ref{sec:massterms}, the quantity $m_{11}^2$ gaps the fermionic spectrum
by a complex conjugate pair of offsets from zero $p^2=\pm i m_{11}^2$. 

In principle, the mean-field analysis also gives access to the curvature of the
effective potential at the minimum. From \Eqref{eq:Veff11v}, we obtain
\begin{equation}
V''(\phi_0=v) = \frac{2 \Nf}{\pi^2} \barh^2.
\label{eq:g11Vpp}
\end{equation}
where the scale is set purely in terms of the (dimensionful) Yukawa coupling. In
the present setting, this result does not acquire an independent meaning. In
order to interpret \Eqref{eq:g11Vpp} as a mass of a scalar $\sigma$-type
excitation on top of the condensate $v$, we would also need the correspondingly
induced kinetic term for this excitation. For instance, if the fluctuation
induced kinetic term read $S[\sigma] = \int \frac{1}{2} Z_\sigma \partial_\mu
\sigma \partial^\mu \sigma$ with a wave function renormalization $Z_\sigma$, the
result for the scalar excitation would be $m_\sigma^2 = \frac{2\Nf}{\pi^2}
\frac{\barh^2}{Z_\sigma}$. 

Finally, as a self-consistency check, we can derive the $\beta$ function for the
scale-dependent coupling $\lambda(\mu)=\barh^2/m^2(\mu)$ within mean-field
theory using \Eqref{eq:renmass11}, yielding 
\begin{equation}
\mu \frac{\partial}{\partial\mu} \lambda(\mu) = - \frac{4N_f}{\pi^2}\lambda^2(\mu),
\label{eq:betag11}
\end{equation}
in agreement with previous results. For the $\g_{11}$ model, the mean-field
computation proceeds fully in the asymptotically free $\lambda>0$ branch of the
model. 

\subsection{$\g_{01}$ model}

Let us now study the fourth model with a scalar self-interaction channel on the
mean-field level. The computation is interesting, since it requires slightly
different techniques. The action reads in the Minkowskian domain 
\begin{equation}\label{eqn:S01M}
S=\int d^4x \left[- \bar\psi G_{\mu\nu}\partial^\mu \partial^\nu \psi - 
\frac{\bar\lambda}{2} (\bar\psi \g_{01} \psi)^2\right].
\end{equation}
Bilinearizing this action by an HS transformation, we arrive at
\begin{equation}\label{eqn:SFB01}
S_{\text{FB}}=\int d^4x \left[- \bar\psi G_{\mu\nu}\partial^\mu \partial^\nu \psi 
+ i \barh \phi \bar\psi \g_{01} \psi -\frac12 \bar{m}^2 \phi^2\right],
\end{equation}
where the $i$ in front of the Yukawa term renders the action real as  in
\Eqref{eqn:SFBg10}, c.f. also \Eqref{eq:mass01}.
The two actions are equivalent if the matching condition \eqref{eq:matching11}
is satisfied. The mean-field approximation thus gives us information about the
asymptotically free $\lambda>0$ branch.

As before, we write the mean-field quantum contribution to the effective action
in terms of the fermion determinant, 
\begin{equation}
\Gamma_{1\ell} = -i \ln\det [-G_{\mu\nu} \partial^\mu \partial^\nu 
+i \barh \phi_0 \g_{01}].
\end{equation}
We have not found a way to rewrite the determinant in terms of scalar squares of
the involved operators as no obvious $\g_{A}$-hermiticity property for a
suitable value of $A$ appears to be available. Hence, we keep the nontrivial
spin structure for the propertime representation of the $\ln\det$. Assuming
$\phi_0=$const., and using that $(-G_{\mu\nu}\partial^\mu\partial^\nu)^{-1}=
\frac{-G_{\mu\nu}\partial^\mu\partial^\nu}{(-\partial^2)^2}$, we can write the
vacuum-subtracted expression as
\begin{equation}
    \Gamma_{1\ell} = -i \Tr \ln \left( \mathds{1} 
    + i \frac{\barh \phi_0}{(-\partial^2)} 
    \frac{(-G_{\mu\nu}\partial^\mu \partial^\nu)}{(-\partial^2)} \g_{01} \right),
\end{equation}
where $\ln\det=\Tr\ln$ has been used. Having in mind that the
coordinate/momentum trace will ultimately be performed in Euclidean momentum
space, we note that the involved operators possess simple hermiticity
properties. The latter imply that the eigenvalues of the total operator in
parentheses in the Euclidean must be of the form $1+ix$ with $x\in \mathds{R}$.
Hence, we can use the standard propertime representation of the logarithm such
that we obtain in momentum space
\begin{equation}
    \Gamma_{1\ell} = -i \Tr \int  \frac{dt}{t} e^{- t}
    \left( \mathds{1}-e^{-i \frac{\barh\phi_0}{p^2} 
    \frac{G_{\mu\nu}p^\mu p^\nu}{p^2} \g_{01}t}\right)
    \label{eq:Gam1ell01}
\end{equation}
Since $\left( \frac{G_{\mu\nu}p^\mu p^\nu}{p^2} \g_{01}\right)^2=\mathds{1}$,
the last exponential in \Eqref{eq:Gam1ell01} can be decomposed as
\begin{equation}
    e^{-i \frac{\barh\phi_0}{p^2} \frac{G_{\mu\nu}p^\mu p^\nu}{p^2} \g_{01}t} 
    = \mathds{1} \cos \frac{\barh\phi_0}{p^2} t 
    -i \frac{G_{\mu\nu}p^\mu p^\nu}{p^2} \g_{01} \sin  \frac{\barh\phi_0}{p^2} t. 
    \label{eq:cosformula} 
\end{equation}
The contribution proportional to $\frac{G_{\mu\nu}p^\mu p^\nu}{p^2}$ vanishes,
since the functional trace, i.e., the momentum integral requires the Lorentz
tensor stucture to be proportional to the metric $\int_p p^{\mu} p^\nu f(p^2)
\sim g^{\mu\nu}$; however, the Lorentz trace of the $G_{\mu\nu}$ vanishes,
$G^{\mu}{}_\mu=0$. The trace in spinor and flavor space thus becomes trivial.
Next, we can Wick rotate the momentum-space variables to the Euclidean domain,
rescale the propertime $t \to p^2 t$, perform the momentum integral and arrive
at
\begin{equation}
\Gamma_{1\ell} \frac{\Nf d_{\g}}{16 \pi^2} \Omega 
\int_{1/\Lambda^2}^\infty \frac{dt}{t^3} (1- \cos \barh \phi_0 t), 
\label{eq:Gam1ell01b}
\end{equation}
where we have introduced a UV cutoff at the lower bound of the propertime
integral. The integral can be evalutated analytically in terms of cosine
integral functions. Expanding the result for large UV cutoff $\Lambda$ and
dropping the terms that vanish in the limit of  $\Lambda\to\infty$, we obtain
for the effective potential
\begin{eqnarray}
V_{\text{eff}}(\phi_0)\!\!&=&\!\!\phi_0^2 \left\{\frac{\bar{m}^2}{2} 
    - \frac{N_f d_{\g}}{2^6 \pi^2}\barh^2 
    \left[3-2\g-\ln\left(\frac{(\barh\phi_0)^2}{\Lambda^4}\right)\right] 
    \right\}\nonumber\\
&=&\!\!\frac{\phi_0^2}{2}\!\! \left\{{m}^2(\mu) 
    - \frac{N_f d_{\g}}{2^5 \pi^2}\barh^2 \left[3-2\g
    -\ln\left(\!\frac{(\bar h\phi_0)^2}{\mu^4}\!\right)\!\right]\! \right\},\nonumber\\
&&\label{eq:Veff01a}
\end{eqnarray}
where we have introduced the renormalized mass $m(\mu)$ using the same
definition as in \Eqref{eq:renmass11} for the $\g_{11}$ model. As in this
previous model, the effective potential features a nontrivial minimum at
\begin{equation}
    (\barh v)^2 = {\mu^4}\, 
    e^{-\frac{2^5\pi^2}{N_f d_{\g}}\frac{m^2(\mu)}{\barh^2}+2-2\gamma},
    \label{eq:vmin01}
\end{equation}
which is RG invariant, since $\mu \frac{d}{d\mu} v=0$. In terms of this minimum,
the effective potential for the present $\g_{01}$ model can be written in the
identical form of \Eqref{eq:Veff11v} as for the $\g_{11}$ model. Accordingly,
its graph is identical to that shown in Fig.~\ref{fig:YZ}. If the absence of the
fermionic mass term is protected by a suitable symmetry, it is broken by the
ground state $\phi_0=v$ at the mean-field quantum level, inducing a fermion mass
\begin{equation}
    m_{01}^2=\barh v. 
    \label{eq:m01induced}
\end{equation}
Similar to the $\g_{11}$ model, the model exhibits dimensional transmutation and
induces a gap in the fermion spectrum in terms of a complex conjugate pair
of offsets from zero $p^2=\pm i m_{01}^2$, as below \Eqref{eq:m01poles}.
Analogously, the model allows for massive $\sigma$-type scalar excitations on
top of the ground state as is indicated by the curvature of the effective
potential, cf. \Eqref{eq:g11Vpp}. Of course, the present mean-field analysis
also passes the self-consistency check in terms of the $\beta$ function for the
scale-dependent coupling $\lambda(\mu)=\barh^2/m^2(\mu)$ also yielding
\Eqref{eq:betag11} as in all other cases.

\subsection{LNJL model}

Let us finally take a look at the mean-field result for the Luttinger-fermionic
version of the NJL model, featuring a continuous axial $\U(1)_{10}$ symmetry.
The action in Minkoski space reads
\begin{equation}\label{eqn:SNJL}
 S=\int d^4x \left[-Z_\psi\bar\psi G_{\mu\nu}\partial^\mu \partial^\nu \psi -
\frac{\bar\lambda}{2}[(\bar\psi\psi)^2-(\bar\psi \gamma_{10}\psi)^2]\right].
\end{equation}
Both interaction channels can be bilinearized by an HS transformation involving
this time a complex  massive scalar field, yielding the partially bosonized
version of action
\begin{eqnarray}\label{eqn:SFBNJL}
S_{\text{FB}}=\int d^4x \biggl[&-Z_\psi \bar\psi G_{\mu\nu}\partial^\mu \partial^\nu \psi 
+ \barh \phi \bar\psi \left(\frac{1-\g_{10}}{2}\right) \psi \nonumber\\
&+ \barh^* \phi^* \bar\psi \left( \frac{1+\g_{10}}{2} \right)  \psi - \bar{m}^2 \phi^* \phi\biggr].
\end{eqnarray}
Here we allow for a complex Yukawa-like coupling $\barh$ for generality. Reality
of the action \eqref{eqn:SFBNJL} is again manifest, since the two fermion-boson
interaction terms are complex conjugate to one another. The matching condition
for the two models to be identical now reads
\begin{equation}
\bar\lambda = - \frac{|\barh|^2}{2\bar{m}^2},
\label{eq:HSmatchingNJL}
\end{equation}
where the additional factor of two in the denominator, e.g., in comparison to
\Eqref{eqn:HSrelation}, is a consequence of the complex-field normalization. As
for the LGN or the $\g_{10}$ model, the HS transformation can be meaningfully
performed only for negative values of the coupling $\bar\lambda$. Incidentally,
we observe from \Eqref{eq:HSmatchingNJL} that $\barh$ could have been chosen
real from the outset. Also, any complex phase of $\barh$ can be compensated by a
global phase rotation of the field $\phi$.

For the choice of the ground state in a mean-field computation, we also have a
free phase parameter to choose. Therefore, we assume $\barh$ as well as
$\phi=\phi_0\in \mathbb{R}$ as real without loss of generality, and compute the
one-loop contribution $\Gamma_{1\ell}$ to the effective action for
$\phi_0=$const. In fact, we again arrive at the same result as for the LGN model
in \Eqref{eq:Gamma1ellLGN}.  

Also, all other conclusions such as the occurrence of an imaginary part of the
effective action for finite $\phi_0$ as a result of the tachyonic quantum modes,
and $\phi_0=0$ being the only equilibrium state of the effective action are
essentially the same as for the LGN or the $\g_{10}$ model. For completeness, we
state the resulting renormalized effective potential accounting for the factor
of two difference of the field normalization
\begin{equation} \label{eqn:veffnjlmu}
    V_{\text{eff}}=|\phi_0|^2 \left\{m^2(\mu) + \frac{N_f d_{\g}}{2^6 \pi^2}|\barh|^2 
    \left[1-\g-\ln\left(\frac{|\barh\phi_0|^2}{\mu^4}\right)\right] \right\}.
\end{equation}
where the renormalized mass $m^2(\mu)$ has been defined as in
\Eqref{eq:GNrenmass}. We also have written the effective potential such that it
is valid for any constant complex field value $\phi_0\in\mathbb{C}$. We conclude
that the LNJL model does not exhibit a gap formation on the negative
$\bar\lambda$ branch that is accessible by the standard HS transformation. 

We conclude this section by mentioning that the mean-field computation also
gives access to the running of the renormalized LNJL coupling defined by
$\lambda(\mu)=-|\barh|^2/(2m^2(\mu))$, yielding the large-$\Nf$ beta function
\Eqref{eqn:betafunctionFRG} as expected. 

\subsection{Summary of mean-field results}

Let us summarize our findings for all the scalar self-interacting models at
mean-field level in Table \ref{table:table}. All models that we considered
exhibit asymptotic freedom for positive values of the coupling $\lambda$; in
fact, the sign conventions of the interaction terms have deliberately been
chosen such that the one-loop RG flows exhibit the same sign. The mean-field
analysis has been performed with the aid of the Hubbard-Stratonovich
transformation in the positive (asymptotically free) branch of the coupling
$\lambda$ for the $\g_{11}$ and $\g_{01}$ models, and in the negative
(non-asymptotically free) branch for the LGN, $\g_{10}$ and LNJL models. 

In the mean-field approximation, only the $\g_{11}$ and $\g_{01}$ models feature
the formation of a nonzero condensate in the effective potential and thereby gap
formation in the fermion spectrum on the coupling branch accessible by the HS
transformation. The corresponding gaps, however, do not correspond to a
conventional real mass term, but to a complex pair of offsets in imaginary
direction in the complex $p^2$ plane. 

The other models do not undergo gap formation in the mean-field approximation,
i.e. the fermions remain massless and the effective potential exhibits a global
minimum at $\phi_0=0$. While the HS transformation gives access only to the
branch which is not asymptotically free, condensation or gap formation is not
observed at all even for arbitrarily large (negative) bare coupling values. In a
sense, this result can be interpreted as a self-consistent behavior of these
models: if gap formation had occurred, the fermionic spectrum would have
featured tachyonic modes. At the same time, the effective
action would have acquired imaginary parts indicating the instability of such a
ground state.

\begin{table}
\begin{center}
\begin{tabular}{ c|C{1.8cm}|C{1.8cm}|C{1.8cm}|C{1.8cm} } 
Model & Asymptotic freedom & Mean-field analysis & Gap formation in MF & Fermionic spectrum \\ 
 \hline
 LGN & $\lambda>0$ & $\lambda<0$ & no & massless\\ 
 $\g_{10}$ & $\lambda>0$ & $\lambda<0$ & no & massless\\ 
 $\g_{11}$ & $\lambda>0$ & $\lambda>0$ & yes & complex gap\\ 
 $\g_{01}$ & $\lambda>0$ & $\lambda>0$ & yes & complex gap\\ 
 LNJL & $\lambda>0$ & $\lambda<0$ & no & massless\\ 
\end{tabular}
\end{center}
\caption{Summary of mean-field level results for all fermionic models studied in this work. While all models feature an asymptotically free branch for $\lambda>0$, the matching condition required by the standard Hubbard-Stratonovich transformation gives access to the mean-field analysis only for a specific branch. For the two models ($\g_{11}$ and $\g_{01}$) for which we can analyse the long-range behavior of the asymptotically free branch, we find condensate and gap formation at mean-field level and a gap in the complexified fermion spectrum.} \label{table:table}
\end{table}

\section{Analytic structure of gapped propagators}
\label{sec:propagators}

As discussed in Sec.~\ref{sec:massterms}, the classical equations of motion of
the free theory admits tachyonic solutions for the case of the standard mass
term and the $m_{10}$ mass. In the interacting cases of the LGN, the $\g_{10}$,
and the LNJL model, these modes have the potential to trigger an instability
associated with an imaginary part of the effective action, if the ground state
developed a fermionic condensate. At mean-field level and for the negative
coupling branch, this, however, did not happen, since the ground state remains
trivial. 

Nevertheless, the potential presence of an instability may be viewed as a
manifestation of Ostrogradsky's theorem stating that Hamiltonians of
higher-derivative theories are unbounded from below \cite{Ostrogradsky:1850fid}.
On the quantum level, higher-derivative theories generically go along with
\textit{ghosts}, i.e., states manifesting as poles in the propagator with
negative spectral weight
\cite{Pais:1950za,Lee:1970iw,Stelle:1976gc,Grinstein:2007mp,Woodard:2015zca}. In
the sense of the Lehmann-Källen spectral representation, such states do not
allow for a probability interpretation potentially invalidating the existence of
an $S$ matrix and thus the validity of such theories as quantum field theories.
Despite these serious issues at least for a perturbative construction, many
concrete proposals have been made to deal with ghosts in a quantized fundamental
theory
\cite{Lee:1970iw,Narnhofer:1978sw,Hawking:2001yt,Bender:2007wu,Grinstein:2007mp,Garriga:2012pk,Salvio:2014soa,Smilga:2017arl,Becker:2017tcx,Anselmi:2018kgz,Gross:2020tph,Donoghue:2021eto,Platania:2019qvo};
moreover theories with ghosts can be meaningfully discussed within effective
field theory, e.g., in cases where the time scale of the instability is large
compared to other time scales of interest; see
\cite{Caldwell:1999ew,Cline:2003gs,Cline:2023hfw,Cline:2024zhs} for applications
in cosmology. 

\begin{figure}
\centering
\begin{tabular}{|c|c|}
    \hline
         \includegraphics[width=0.224\textwidth]{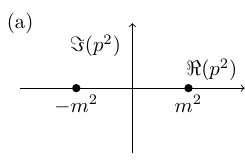} &  \includegraphics[width=0.224\textwidth]{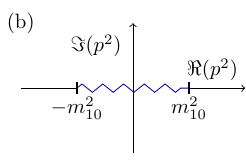} \\ 
       \hline
        \includegraphics[width=0.224\textwidth]{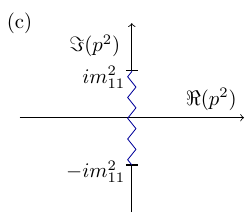} &  \includegraphics[width=0.224\textwidth]{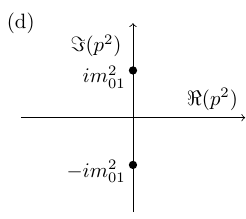}\\
     \hline
    \end{tabular}
    \caption{Analytic structure of the propagators for the free
    theory with (a) a standard mass $m$, (b) an $m_{10}$ mass, (c) an $m_{11}$
    mass, (d) an $m_{01}$ mass depicted in the complex $p^2$ momentum plane.
    Black dots represent the position of simple poles in the eigenvalues of the propagator, zigzag lines denote a branch cut.} 
    \label{tab:polestructure}
\end{figure}

In order to study the possible occurrence of ghosts, let us determine the
analytic structure of the propagators of the various free theories in the
complex $p^2$ momentum plane. We start with the free theory with a standard mass
term \Eqref{eq:standardmass}. The Minkowski space propagator $S(p)$ is related
to the inverse Hessian of the action by
\begin{equation}
-i S(p)= \frac{1}{G_{\mu\nu}p^\mu p^\nu - m^2} 
= \frac{1}{p^4 -m^4} \big( G_{\mu\nu}p^\mu p^\nu + m^2\big),
\label{eq:standardmassprop}
\end{equation}
which is, of course, matrix-valued in spinor space. Using, for instance, the
explicit representation of the Abrikosov algebra given in App.~\ref{sec:AppA},
we can determine the eigenvalues of the propagator:
\begin{equation}
\mathrm{eig}\big( -i S(p)\big) 
= \left\{ \frac{1}{p^2-m^2}\, \big[\mathrm{deg.} 16 \big], 
 \frac{-1}{p^2+m^2}\, \big[\mathrm{deg.} 16 \big] \right\},
\label{eq:standardmasseig}
\end{equation}
where we find that the two different eigenvalues occur with a degree of
degeneracy of 16 each. From \Eqref{eq:standardmasseig}, we can read off that the
propagator features 16 poles in the complex $p^2$ plane that correspond to a
standard massive dispersion relation $p^2=m^2$, and 16 poles corresponding to
tachyonic states with $p^2=-m^2$, cf. Fig.~\ref{tab:polestructure} (a). 
Moreover, the latter are, in fact, ghost poles
as their residue is negative. Of course, causality requirements for the
propagator may be implemented by suitable $i\epsilon$ prescriptions; however,
the details are not relevant for the present discussion.

We conclude that the free theory with a standard mass term indeed exhibits the
properties that are generically expected from a higher-derivative theory: it
features tachyons and ghosts. For the present case, the tachyonic and the ghost
states are identical; these properties are not necessarily linked, a
counterexample is, e.g., given by certain versions of quadratic gravity
\cite{Stelle:1976gc,Buccio:2024hys}. 

As a second example, let us consider the free theory with a $\g_{10}$ mass term
of \Eqref{eq:mass10}. Here, the propagator is given by
\begin{eqnarray}
    -i S(p)&=& \frac{1}{G_{\mu\nu}p^\mu p^\nu - i m_{10}^2 \g_{10}}\nonumber\\ 
    &=& \frac{1}{p^4 -m_{10}^4} \big( G_{\mu\nu}p^\mu p^\nu - i m_{10}^2 \g_{10}\big),
    \label{eq:prop10}
\end{eqnarray}
with eigenvalue spectrum
\begin{eqnarray}
    \mathrm{eig}\big( -i S(p)\big) 
    &=& \left\{ \frac{1}{\sqrt{(p^2-m_{10}^2)(p^2+m_{10}^2)}}\, \big[\mathrm{deg.} 16 \big],\right. \nonumber\\ 
    &&\left. - \frac{1}{\sqrt{(p^2-m_{10}^2)(p^2+m_{10}^2)}}\, \big[\mathrm{deg.} 16 \big] \right\}.\nonumber\\
    &&
    \label{eq:eig10}
\end{eqnarray}
We observe that the propagator has square root singularities at $p^2=\pm
m_{10}^2$ instead of simple poles. This implies that there is a branch cut in
the complex $p^2$ plane. Choosing the cut to lie at negative values of the
radicand, the branch cut extends from $p^2=-m_{10}^2$ to $p^2= m_{10}^2$ along
the real axis, see Fig.~\ref{tab:polestructure} (b). Half of the modes comes
with a minus sign such that we rediscover the ghost modes in the massless limit
$m_{10}^2\to0$ as expected. However, there is no straightforward Lehmann-Källen
spectral representation of the propagator for finite $m_{10}^2$ and thus no
immediate probability interpretation in terms of asymptotic states. 

Let us also study the propagator for the free theory including a $\g_{11}$ mass term, cf. \Eqref{eq:mass11},
\begin{eqnarray}
    -i S(p)&=& \frac{1}{G_{\mu\nu}p^\mu p^\nu - m_{11}^2 \g_{11}}\nonumber\\ 
    &=& \frac{1}{p^4 +m_{11}^4} \big( G_{\mu\nu}p^\mu p^\nu - m_{11}^2 \g_{11}\big).
    \label{eq:prop11}
\end{eqnarray}
Now, the eigenvalue spectrum reads
\begin{eqnarray}
    \mathrm{eig}\big( -i S(p)\big) 
    &=& \left\{ \frac{1}{\sqrt{(p^2-im_{11}^2)(p^2+im_{11}^2)}}\, \big[\mathrm{deg.} 16 \big],\right. \nonumber\\ 
    &&\left. - \frac{1}{\sqrt{(p^2-im_{11}^2)(p^2+im_{10}^2)}}\, \big[\mathrm{deg.} 16 \big] \right\}.\nonumber\\
    &&
    \label{eq:eig11}
\end{eqnarray}
The propagator again supports a square-root type branch cut in the complex $p^2$
plane, this time ranging from $p^2 = -i m_{11}^2$ to $p^2 = i m_{11}^2$ along
the imaginary $p^2$ axis, cf. Fig.~\ref{tab:polestructure} (c). Also in this
case, we observe the absence of a conventional spectral representation thus
losing the interpretation of some of the modes as ghosts.

A fourth interesting example is given by the free theory with a $\g_{01}$ mass,
cf.~\Eqref{eq:mass01} with the propagator
\begin{eqnarray}
    -i S(p)
    &=& \frac{1}{p^4 +m_{01}^4} \big( G_{\mu\nu}p^\mu p^\nu + i m_{01}^2 \g_{01}\big),
    \label{eq:prop01}
\end{eqnarray}
yielding a slightly more intricate eigenvalue spectrum
\begin{eqnarray}
    \mathrm{eig}\big( -i S(p)\big) 
    &=& \left\{ \frac{1}{p^2-im_{01}^2}\, \big[\mathrm{deg.} 8 \big],
    \frac{1}{p^2+im_{01}^2}\, \big[\mathrm{deg.} 8 \big], \right. \nonumber\\ 
    &&\left.   \frac{-1}{p^2-im_{01}^2}\, \big[\mathrm{deg.} 8 \big],
    \frac{-1}{p^2+im_{01}^2}\, \big[\mathrm{deg.} 8 \big]\right\}.\nonumber\\
    &&
    \label{eq:eig01}
\end{eqnarray}
For this mass term, we observe the naively expected simple poles on the imaginary axis at $p^2=\pm i m_{01}^2$. Half of the modes seems to have a ghost-type residue. However, since the poles are off the real axis, there is no conventional spectral representation and thus no straightforward probability interpretation. 

At this point, we conclude that a naively expected link between a higher-derivative theory, the occurrence of ghosts and an inferred break-down of a consistent quantum field theory does not hold in general in theories with relativistic Luttinger fermions. While this link appears to be present in the case of theories with a standard mass term, where we do find tachyonic ghosts, all other mass terms do not give rise to a spectral representation. Therefore, we have no reason to infer that these theories feature ghost states.

Based on these observations read together with the findings of the previous
sections, our interpretation at present is as follows: 

(a) the tachyonic ghost states of relativistic Luttinger fermions with a
standard mass term inhibit a straightforward perturbative construction of
observables such as those derived from an $S$ matrix. This is also reflected in
our mean-field approach for the LGN or LNJL model by the potential occurrence of
an imaginary part of the effective action as a consequence of tachyonic
instabilites. Of course, this does not exclude the possibility that a successful
quantization may be possible along the lines suggested for other
higher-derivative theories. 

(b) While we cannot make a statement about the possible (in-)existence of
perturbative $S$-matrix based observables for Luttinger fermions with $m_{10}^2$
masses due to the lack of a spectral representation, the occurrence of tachyonic
modes in the mean-field approach to the $\g_{10}$ model suggests that such
degrees of freedom lead to similar problems as those with the standard mass
terms and tachyonic ghosts. 

(c) For the models with $m_{11}^2$ or $m_{01}^2$ mass terms, we have observed a
perfectly stable and consistent mean-field description for the corresponding
$\g_{11}$ and $\g_{01}$ models going along with the absence of tachyonic mass
poles. Also, we do not find ghost states in the sense of conventional mass poles
with negative residue. From this perspective, we do not see any reason based on
our analysis why such theories should not be consistent. 

On the other hand, our results suggest that relativistic Luttinger fermions with
$m_{11}^2$ or $m_{01}^2$ mass terms do not have a conventional Lehmann-Källen
spectral representation and thus no conventional LSZ construction of the $S$
matrix. Our preliminary interpretation of this finding is that such relativistic
Luttinger fermions do not exist in the sense of asymptotic states. In fact,
propagators with complex poles have been intensely discussed in the literature
of the strong interactions where the fundamental variables of the QCD action,
quarks and gluons, are not expected to exist as asymptotic states
\cite{Stingl:1985hx,Stingl:1994nk,Alkofer:2000wg,Fischer:2003rp,Alkofer:2003jj,Cucchieri:2004mf,Dudal:2008sp,Cucchieri:2011ig,Hayashi:2018giz,Binosi:2019ecz,Li:2019hyv,Fischer:2020xnb,Huber:2020keu,Horak:2022myj,Braun:2022mgx,Horak:2021syv,Horak:2023xfb}.
Nevertheless, the long-range physics can be described by asymptotic (bound)
states such as hadrons in strong-interaction physics or a composite
$\sigma$-type excitation in the present case of the $\g_{11}$ or $\g_{01}$
model.

\section{Conclusions}
\label{sec:conc}

We have introduced and investigated a number of self-interacting quantum field
theories with relativistic Luttinger fermions as fundamental degrees of freedom.
Concentrating on models with scalar interaction channels, we find that each one
features a coupling branch which is asymptotically free in four dimensional
spacetime. While we have worked at a large number of flavors where the
restriction to the single scalar interaction channels is justified and
quantitatively controlled, we expect our results on asymptotic freedom to
generalize to full Fierz-complete local interaction bases, as has been shown in
\cite{Gies:2023cnd} for the LGN model. 

We provide large-$\Nf$ exact results for the mean-field effective potential for
each model, identifying two models that are high-energy complete and undergo
dimensional transmutation with a corresponding condensate and gap formation at
low energies. Both models, the $\g_{11}$ and $\g_{01}$ model, do not exhibit any
sign of instability at the present level of investigation, as might naively be
expected for a higher-derivative theory. The reason for this lies in the fact
that the generated mass term does not induce a tachyonic mass pole. This is
corroborated by our study of the analytic structure of the propagators for the
various versions of the massive theories. 

It is interesting to observe that mass generation and gap formation does not
happen at mean-field level for those theories where these masses would lead to
tachyonic instabilities. However, since the mean-field analysis is confined to a
specific coupling branch, other methods are needed to analyse the more
interesting asymptotically free branch for those models. This should be possible
with suitable techniques that allow for the inclusion of a more general
bare potential of the scalar field such as the functional RG, see
\cite{Braun:2011pp,Braun:2011pp,Jakovac:2015iqa,Cresswell-Hogg:2022lgg,Cresswell-Hogg:2022lez,Cresswell-Hogg:2023hdg},
or methods based on the gap equation. 

These methods could also provide access to the spectrum of excitations above the
non-trivial ground state in the stable and high-energy complete models. In the
present cases of the $\g_{11}$ and $\g_{01}$ models, we expect the existence of
a light $\sigma$ mode. This would constitute an example of a UV-complete model
with only marginal couplings that entails naturally light scalar long-range
degrees of freedom in four-dimensional spacetime. Corresponding investigations
are underway.

\section*{Acknowledgments}

We thank Gerald Dunne, Philip Heinzel, \'Alvaro Pastor-Guti\'errez and Richard Schmieden for valuable discussions, and Philip Heinzel for collaboration at initial
stages of the project.  
This work has been funded by the Deutsche 
Forschungsgemeinschaft (DFG) under Grant No. 406116891 
within the Research Training Group RTG 2522/1.

\appendix
\setcounter{equation}{0}
\setcounter{figure}{0}
\setcounter{table}{0}
\makeatletter
\renewcommand{\theequation}{A\arabic{equation}}
\renewcommand{\thefigure}{A\arabic{figure}}
\setcounter{secnumdepth}{3}

\section{Relativistic Abrikosov algebra}
\label{sec:AppA}

In order to make the paper self-contained, let us summarize a few aspects of the
relativistic version of the Abrikosov algebra \cite{Abrikosov:1974a} in
\Eqref{eq:AbrikosovA} as derived in \cite{Gies:2023cnd}. 

Whenever needed, we work with the explicit representation of the $G_{\mu\nu}$
matrices in terms of elements of a Euclidean Dirac algebra
$\{\gamma_A,\gamma_B\} = 2 \delta_{AB}$,
\begin{eqnarray}
 G_{0i}&=&i \sqrt{\frac{2}{3}} \gamma_{A=i}, \quad i=1,2,3, \nonumber\\
 G_{12}&=& \sqrt{\frac{2}{3}} \gamma_4, \quad G_{23} = \sqrt{\frac{2}{3}} 
\gamma_5, \quad G_{31}= \sqrt{\frac{2}{3}} \gamma_6, \nonumber\\
G_{00}&=& \gamma_7, \quad G_{11} = \frac{1}{3} \gamma_7 + \frac{2\sqrt{2}}{3} 
\gamma_8,\label{eq:Grep}\\
G_{22}&=& \frac{1}{3} \gamma_7 - \frac{\sqrt{2}}{3} \gamma_8 + 
\sqrt{\frac{2}{3}} \gamma_9, \nonumber\\
G_{33}&=& \frac{1}{3} \gamma_7 - \frac{\sqrt{2}}{3} \gamma_8 - 
\sqrt{\frac{2}{3}} \gamma_9. \nonumber
\end{eqnarray}
This representation can be viewed as an appropriate Wick rotation of   
the one constructed for $d=4$ Euclidean dimensions in \cite{Janssen:2015xga}. It
is straightforward to check that this representation satisfies
\Eqref{eq:AbrikosovA}. Whereas 9 elements $\gamma_{1,\dots,9}$ were sufficient
to satisfy the relativistic Abrikosov algebra, the reality conditions of the
action demand for another anti-commuting element for the construction of a spin
metric $h$. This requires a $d_\gamma=32$ dimensional representation for the
Euclidean Dirac algebra (and correspondingly of the Abrikosov algebra) and thus
in total with $d_e=11$ anticommuting elements $\gamma_A$, with $A=1,\dots, 11$. 

While we use $\gamma_{10}$ for the construction of the spin metric, cf.
\Eqref{eq:spinmetric}, both additional elements $\gamma_{10}$ and $\gamma_{11}$
can serve for the construction of additional scalar bilinears and interactions.
Alternatively, we could choose a different spin metric, e.g., 
\begin{equation}
\tilde{h}=\gamma_1\gamma_2\gamma_3\gamma_{11},
\label{eq:htildeh}
\end{equation}
or, more generally, a linear combination $h'=\alpha h + \beta \tilde{h}$ with
$\alpha^2+\beta^2=1$ and $\alpha,\beta\in \mathbb{R}$ as the spin metric. This
would induce a corresponding rotation of the interaction channels discussed in
the main text, but the overall structure of different mass terms, interaction
channels and the existence of an axial symmetry of the kinetic term would remain
the same.

\section{Euclidean conventions}
\label{sec:AppC}

For the Wilsonian renormalization group analysis in the main text, it is useful
to have a manifestly Euclidean formulation of the models studied in the present
work. For this, we need a Euclidean version of the Abrikosov algebra, 
\begin{equation}
    \left.\{G_{\mu\nu}, G_{\kappa\lambda}\}\right|_E = - \frac{2}{d-1} 
   \delta_{\mu\nu}\delta_{\kappa\lambda}+\frac{d}{d-1} (\delta_{\mu\kappa} \delta_{\nu\lambda} + 
   \delta_{\mu\lambda}\delta_{\nu\kappa}),
   \label{eq:AbrikosovAEU}
\end{equation}
where Minkowski metric factors on the right-hand side are replaced by Kronecker
symbols. For generality, we work in $d$-dimensional spacetime here. We also introduce $\tau$ as Euclidean time direction, related to the
Minkowskian time $t$ by a Wick rotation $\tau=it$. The reality condition of the
Minkowskian action can then be rephrased for the Euclidean Lagrangian in terms
of Osterwalder-Schrader (OS) reflection positivity \cite{Osterwalder:1973dx},
\begin{equation}
\mathcal{L}_E^*=\hat{\mathcal{L}}_E,
\end{equation}
where $\hat{\mathcal{L}}_E$ arises from ${\mathcal{L}}_E$ by replacing the
coordinates $x=(\tau,\vec{x})$ with $\hat{x}=(-\tau,\vec{x})$. In other words,
in addition to complex conjugating the operator building blocks, we also need to
flip the sign of the Euclidean time. Let us first check OS reflection positivity
for the simple mass term. We extend the fields to the Euclidean domain
$\psi=\psi(\tau,\vec{x})$ and look at the complex conjugate of the spinor
bilinear
\begin{equation}
[(\bar\psi\psi)(x)]^*=(\psi^\dagger h^\dagger \psi)(\hat{x}).
\end{equation}
A simple choice to implement reflection positivity here is to define the
Euclidean spin metric to be equivalent to the hermitean Minkowskian one,
$h_E=h\equiv h^\dagger$, also preserving the definition $\bar\psi = \psi^\dagger
h$.

Prior to looking at the kinetic term, let us establish the connection between
the Euclidean $G_{\mu\nu}$ matrices, satisfying \Eqref{eq:AbrikosovAEU}, and
their Minkowskian counterpart. From the algebra \eqref{eq:AbrikosovAEU}, all the
$G_{\mu\nu,E}$ matrices can be chosen hermitean in the Euclidean domain.
Furthermore, by comparing \eqref{eq:AbrikosovA} and \eqref{eq:AbrikosovAEU}, we
can find the explicit relation of the matrices in the Euclidean and Minkowskian
domain. For example, by looking at the anticommutator
\begin{equation}
\left.\{G_{00}, G_{00}\}\right|_M =2=\left.\{G_{00}, G_{00}\}\right|_E,
\end{equation}
we infer that the choice $G_{00,E}\equiv G_{00,M}$ is a valid option.
Next, since
\begin{equation}
\left.\{G_{00},G_{\underline{i}\underline{i}}\}\right|_M = 2/(d-1)
=-\left.\{G_{00},G_{\underline{i}\underline{i}}\}\right|_E,
\end{equation}
we can take $G_{ii,E}\equiv - G_{ii,M}$.
Another non-trivial anti-commutator is
\begin{equation}
\left.\{G_{0\underline{i}},G_{0\underline{i}}\}\right|_M = -d/(d-1)
=-\left.\{G_{0\underline{i}},G_{0\underline{i}}\}\right|_E,
\end{equation}
from which we read off the two possibilities $\pm i G_{0i,E}\equiv G_{0i,M}$.
Lastly, noticing that
\begin{equation}
\left.\{G_{\underline{i}\underline{j}},G_{\underline{i}\underline{j}}\}\right|_M = d/(d-1)
=\left.\{G_{\underline{i}\underline{j}},G_{\underline{i}\underline{j}}\}\right|_E,
\end{equation}
we get $\pm G_{ij,E}\equiv G_{ij,M}$. The last two ambiguities can be resolved
by looking at the Luttinger kinetic operator
$G_{\mu\nu}\partial^\mu\partial^\nu$. For the $00$ and $ii$ components, we have
\begin{equation}\begin{split}\label{eqn:LuttOpEM}
\left.G_{00}\partial^0\partial^0\right|_M &
= \left.G_{00}\right|_E \frac{1}{(-i)^2}\frac{\partial}{\partial\tau}\frac{\partial}{\partial\tau}= - \left.G_{00}\partial^0\partial^0\right|_E, \\
\left.G_{\underline{i}\underline{i}}\partial^{\underline{i}}\partial^{\underline{i}}\right|_M &
= -\left.G_{\underline{i}\underline{i}}\right|_E (-1)^2 \left.\frac{\partial}{\partial x^{\underline{i}}}\frac{\partial}{\partial x^{\underline{i}}}\right|_M
= - \left.G_{\underline{i}\underline{i}}\partial^{\underline{i}}\partial^{\underline{i}}\right|_E.
\end{split}\end{equation}
Thus, for consistency with equation \eqref{eqn:LuttOpEM}, the choice of the
minus sign also for the $ij$ and $0i$ components, fixes the signs of the
remaining $G_{\mu\nu}$ matrices by $G_{0i,E}\equiv i G_{0i,M}$ and
$G_{ij,E}\equiv - G_{ij,M}$. Ultimately, we obtain for the Luttinger kinetic
term
\begin{equation}\label{eqn:euclkin}
\left. \bar\psi G_{\mu\nu} i \partial^\mu i \partial^\nu \psi \right|_M 
= \left. - \bar\psi G_{\mu\nu} i \partial^\mu i \partial^\nu \psi \right|_E.
\end{equation}
It is straightforward to check that the kinetic term \eqref{eqn:euclkin} is OS
reflection positive in the Euclidean domain. In order for the weight functions
of the functional integrals to undergo the transition $e^{iS_M}$ to $e^{-S_E}$,
we observe that
\begin{equation}\begin{split}
i S_M &= i \int d^4x (\bar\psi G_{\mu\nu} i \partial^\mu i \partial^\nu \psi 
+ \mathcal{L}_{\text{int}}) \\
&= \int \left.d^4x \right|_E (- \left.\bar\psi G_{\mu\nu} i \partial^\mu i \partial^\nu \psi \right|_E 
+ \mathcal{L}_{\text{int}}) \coloneqq - S_E,
\end{split}\end{equation}
holds for any local non-derivative interaction (or mass) term
$\mathcal{L}_{int}$. We conclude that the Euclidean and Minkowskian actions
differ by a minus sign with respect to the local non-derivative terms
$\mathcal{L}_{int}$, 
\begin{equation}
S_E = \int d^4x ( \bar\psi G_{\mu\nu} i \partial^\mu i \partial^\nu \psi  
- \mathcal{L}_{\text{int}}).
\end{equation}
This sign change applies to all interacting models investigated in the present
work, e.g., involving interaction terms such as $(\bar\psi\psi)^2$, $(\bar\psi
\g_{10}\psi)^2$, $(\bar\psi\g_{11}\psi)^2$, $(\bar\psi\g_{01}\psi)^2$,
$(\bar\psi G_{\mu\nu}\psi)^2$, $\dots$.

\bibliography{bibliography}

\begin{thebibliography}{91}%
\makeatletter
\providecommand \@ifxundefined [1]{%
 \@ifx{#1\undefined}
}%
\providecommand \@ifnum [1]{%
 \ifnum #1\expandafter \@firstoftwo
 \else \expandafter \@secondoftwo
 \fi
}%
\providecommand \@ifx [1]{%
 \ifx #1\expandafter \@firstoftwo
 \else \expandafter \@secondoftwo
 \fi
}%
\providecommand \natexlab [1]{#1}%
\providecommand \enquote  [1]{``#1''}%
\providecommand \bibnamefont  [1]{#1}%
\providecommand \bibfnamefont [1]{#1}%
\providecommand \citenamefont [1]{#1}%
\providecommand \href@noop [0]{\@secondoftwo}%
\providecommand \href [0]{\begingroup \@sanitize@url \@href}%
\providecommand \@href[1]{\@@startlink{#1}\@@href}%
\providecommand \@@href[1]{\endgroup#1\@@endlink}%
\providecommand \@sanitize@url [0]{\catcode `\\12\catcode `\$12\catcode
  `\&12\catcode `\#12\catcode `\^12\catcode `\_12\catcode `\%12\relax}%
\providecommand \@@startlink[1]{}%
\providecommand \@@endlink[0]{}%
\providecommand \url  [0]{\begingroup\@sanitize@url \@url }%
\providecommand \@url [1]{\endgroup\@href {#1}{\urlprefix }}%
\providecommand \urlprefix  [0]{URL }%
\providecommand \Eprint [0]{\href }%
\providecommand \doibase [0]{http://dx.doi.org/}%
\providecommand \selectlanguage [0]{\@gobble}%
\providecommand \bibinfo  [0]{\@secondoftwo}%
\providecommand \bibfield  [0]{\@secondoftwo}%
\providecommand \translation [1]{[#1]}%
\providecommand \BibitemOpen [0]{}%
\providecommand \bibitemStop [0]{}%
\providecommand \bibitemNoStop [0]{.\EOS\space}%
\providecommand \EOS [0]{\spacefactor3000\relax}%
\providecommand \BibitemShut  [1]{\csname bibitem#1\endcsname}%
\let\auto@bib@innerbib\@empty
\bibitem [{\citenamefont {Luttinger}(1956)}]{LuttingerPhysRev.102.1030}%
  \BibitemOpen
  \bibfield  {author} {\bibinfo {author} {\bibfnamefont {J.~M.}\ \bibnamefont
  {Luttinger}},\ }\href {\doibase 10.1103/PhysRev.102.1030} {\bibfield
  {journal} {\bibinfo  {journal} {Phys. Rev.}\ }\textbf {\bibinfo {volume}
  {102}},\ \bibinfo {pages} {1030} (\bibinfo {year} {1956})}\BibitemShut
  {NoStop}%
\bibitem [{\citenamefont {Abrikosov}(1974)}]{Abrikosov:1974a}%
  \BibitemOpen
  \bibfield  {author} {\bibinfo {author} {\bibfnamefont {A.~A.}\ \bibnamefont
  {Abrikosov}},\ }\href
  {http://www.jetp.ras.ru/cgi-bin/e/index/e/39/4/p709?a=list} {\bibfield
  {journal} {\bibinfo  {journal} {Sov. Phys. JETP}\ }\textbf {\bibinfo {volume}
  {39}},\ \bibinfo {pages} {709} (\bibinfo {year} {1974})}\BibitemShut
  {NoStop}%
\bibitem [{\citenamefont {Murakami}\ \emph {et~al.}(2004)\citenamefont
  {Murakami}, \citenamefont {Nagaosa},\ and\ \citenamefont
  {Zhang}}]{Murakami:2004zz}%
  \BibitemOpen
  \bibfield  {author} {\bibinfo {author} {\bibfnamefont {S.}~\bibnamefont
  {Murakami}}, \bibinfo {author} {\bibfnamefont {N.}~\bibnamefont {Nagaosa}}, \
  and\ \bibinfo {author} {\bibfnamefont {S.-C.}\ \bibnamefont {Zhang}},\ }\href
  {\doibase 10.1103/PhysRevB.69.235206} {\bibfield  {journal} {\bibinfo
  {journal} {Phys. Rev. B}\ }\textbf {\bibinfo {volume} {69}},\ \bibinfo
  {pages} {235206} (\bibinfo {year} {2004})},\ \Eprint
  {http://arxiv.org/abs/cond-mat/0310005} {arXiv:cond-mat/0310005} \BibitemShut
  {NoStop}%
\bibitem [{\citenamefont {Moon}\ \emph {et~al.}(2013)\citenamefont {Moon},
  \citenamefont {Xu}, \citenamefont {Kim},\ and\ \citenamefont
  {Balents}}]{Moon:2012rx}%
  \BibitemOpen
  \bibfield  {author} {\bibinfo {author} {\bibfnamefont {E.-G.}\ \bibnamefont
  {Moon}}, \bibinfo {author} {\bibfnamefont {C.}~\bibnamefont {Xu}}, \bibinfo
  {author} {\bibfnamefont {Y.~B.}\ \bibnamefont {Kim}}, \ and\ \bibinfo
  {author} {\bibfnamefont {L.}~\bibnamefont {Balents}},\ }\href {\doibase
  10.1103/PhysRevLett.111.206401} {\bibfield  {journal} {\bibinfo  {journal}
  {Phys. Rev. Lett.}\ }\textbf {\bibinfo {volume} {111}},\ \bibinfo {pages}
  {206401} (\bibinfo {year} {2013})},\ \Eprint {http://arxiv.org/abs/1212.1168}
  {arXiv:1212.1168 [cond-mat.str-el]} \BibitemShut {NoStop}%
\bibitem [{\citenamefont {Savary}\ \emph {et~al.}(2014)\citenamefont {Savary},
  \citenamefont {Moon},\ and\ \citenamefont {Balents}}]{Savary:2014gka}%
  \BibitemOpen
  \bibfield  {author} {\bibinfo {author} {\bibfnamefont {L.}~\bibnamefont
  {Savary}}, \bibinfo {author} {\bibfnamefont {E.-G.}\ \bibnamefont {Moon}}, \
  and\ \bibinfo {author} {\bibfnamefont {L.}~\bibnamefont {Balents}},\ }\href
  {\doibase 10.1103/PhysRevX.4.041027} {\bibfield  {journal} {\bibinfo
  {journal} {Phys. Rev. X}\ }\textbf {\bibinfo {volume} {4}},\ \bibinfo {pages}
  {041027} (\bibinfo {year} {2014})},\ \Eprint {http://arxiv.org/abs/1403.5255}
  {arXiv:1403.5255 [cond-mat.str-el]} \BibitemShut {NoStop}%
\bibitem [{\citenamefont {Herbut}\ and\ \citenamefont
  {Janssen}(2014)}]{Herbut:2014lfa}%
  \BibitemOpen
  \bibfield  {author} {\bibinfo {author} {\bibfnamefont {I.~F.}\ \bibnamefont
  {Herbut}}\ and\ \bibinfo {author} {\bibfnamefont {L.}~\bibnamefont
  {Janssen}},\ }\href {\doibase 10.1103/PhysRevLett.113.106401} {\bibfield
  {journal} {\bibinfo  {journal} {Phys. Rev. Lett.}\ }\textbf {\bibinfo
  {volume} {113}},\ \bibinfo {pages} {106401} (\bibinfo {year} {2014})},\
  \Eprint {http://arxiv.org/abs/1404.5721} {arXiv:1404.5721 [cond-mat.str-el]}
  \BibitemShut {NoStop}%
\bibitem [{\citenamefont {Janssen}\ and\ \citenamefont
  {Herbut}(2015)}]{Janssen:2015xga}%
  \BibitemOpen
  \bibfield  {author} {\bibinfo {author} {\bibfnamefont {L.}~\bibnamefont
  {Janssen}}\ and\ \bibinfo {author} {\bibfnamefont {I.~F.}\ \bibnamefont
  {Herbut}},\ }\href {\doibase 10.1103/PhysRevB.92.045117} {\bibfield
  {journal} {\bibinfo  {journal} {Phys. Rev. B}\ }\textbf {\bibinfo {volume}
  {92}},\ \bibinfo {pages} {045117} (\bibinfo {year} {2015})},\ \Eprint
  {http://arxiv.org/abs/1503.04242} {arXiv:1503.04242 [cond-mat.str-el]}
  \BibitemShut {NoStop}%
\bibitem [{\citenamefont {Janssen}\ and\ \citenamefont
  {Herbut}(2016)}]{Janssen:2015jga}%
  \BibitemOpen
  \bibfield  {author} {\bibinfo {author} {\bibfnamefont {L.}~\bibnamefont
  {Janssen}}\ and\ \bibinfo {author} {\bibfnamefont {I.~F.}\ \bibnamefont
  {Herbut}},\ }\href {\doibase 10.1103/PhysRevB.93.165109} {\bibfield
  {journal} {\bibinfo  {journal} {Phys. Rev. B}\ }\textbf {\bibinfo {volume}
  {93}},\ \bibinfo {pages} {165109} (\bibinfo {year} {2016})},\ \Eprint
  {http://arxiv.org/abs/1509.01737} {arXiv:1509.01737 [cond-mat.str-el]}
  \BibitemShut {NoStop}%
\bibitem [{\citenamefont {Boettcher}\ and\ \citenamefont
  {Herbut}(2016)}]{Boettcher:2016wft}%
  \BibitemOpen
  \bibfield  {author} {\bibinfo {author} {\bibfnamefont {I.}~\bibnamefont
  {Boettcher}}\ and\ \bibinfo {author} {\bibfnamefont {I.~F.}\ \bibnamefont
  {Herbut}},\ }\href {\doibase 10.1103/PhysRevB.93.205138} {\bibfield
  {journal} {\bibinfo  {journal} {Phys. Rev. B}\ }\textbf {\bibinfo {volume}
  {93}},\ \bibinfo {pages} {205138} (\bibinfo {year} {2016})},\ \Eprint
  {http://arxiv.org/abs/1603.00031} {arXiv:1603.00031 [cond-mat.str-el]}
  \BibitemShut {NoStop}%
\bibitem [{\citenamefont {Janssen}\ and\ \citenamefont
  {Herbut}(2017)}]{Janssen:2016xvc}%
  \BibitemOpen
  \bibfield  {author} {\bibinfo {author} {\bibfnamefont {L.}~\bibnamefont
  {Janssen}}\ and\ \bibinfo {author} {\bibfnamefont {I.~F.}\ \bibnamefont
  {Herbut}},\ }\href {\doibase 10.1103/PhysRevB.95.075101} {\bibfield
  {journal} {\bibinfo  {journal} {Phys. Rev. B}\ }\textbf {\bibinfo {volume}
  {95}},\ \bibinfo {pages} {075101} (\bibinfo {year} {2017})},\ \Eprint
  {http://arxiv.org/abs/1611.04594} {arXiv:1611.04594 [cond-mat.str-el]}
  \BibitemShut {NoStop}%
\bibitem [{\citenamefont {Boettcher}\ and\ \citenamefont
  {Herbut}(2017)}]{Boettcher:2016iiv}%
  \BibitemOpen
  \bibfield  {author} {\bibinfo {author} {\bibfnamefont {I.}~\bibnamefont
  {Boettcher}}\ and\ \bibinfo {author} {\bibfnamefont {I.~F.}\ \bibnamefont
  {Herbut}},\ }\href {\doibase 10.1103/PhysRevB.95.075149} {\bibfield
  {journal} {\bibinfo  {journal} {Phys. Rev. B}\ }\textbf {\bibinfo {volume}
  {95}},\ \bibinfo {pages} {075149} (\bibinfo {year} {2017})},\ \Eprint
  {http://arxiv.org/abs/1611.05904} {arXiv:1611.05904 [cond-mat.str-el]}
  \BibitemShut {NoStop}%
\bibitem [{\citenamefont {Ray}\ \emph {et~al.}(2018)\citenamefont {Ray},
  \citenamefont {Vojta},\ and\ \citenamefont {Janssen}}]{Ray:2018gtp}%
  \BibitemOpen
  \bibfield  {author} {\bibinfo {author} {\bibfnamefont {S.}~\bibnamefont
  {Ray}}, \bibinfo {author} {\bibfnamefont {M.}~\bibnamefont {Vojta}}, \ and\
  \bibinfo {author} {\bibfnamefont {L.}~\bibnamefont {Janssen}},\ }\href
  {\doibase 10.1103/PhysRevB.98.245128} {\bibfield  {journal} {\bibinfo
  {journal} {Phys. Rev. B}\ }\textbf {\bibinfo {volume} {98}},\ \bibinfo
  {pages} {245128} (\bibinfo {year} {2018})},\ \Eprint
  {http://arxiv.org/abs/1810.07695} {arXiv:1810.07695 [cond-mat.str-el]}
  \BibitemShut {NoStop}%
\bibitem [{\citenamefont {Ray}\ \emph {et~al.}(2020)\citenamefont {Ray},
  \citenamefont {Vojta},\ and\ \citenamefont {Janssen}}]{Ray:2020mlg}%
  \BibitemOpen
  \bibfield  {author} {\bibinfo {author} {\bibfnamefont {S.}~\bibnamefont
  {Ray}}, \bibinfo {author} {\bibfnamefont {M.}~\bibnamefont {Vojta}}, \ and\
  \bibinfo {author} {\bibfnamefont {L.}~\bibnamefont {Janssen}},\ }\href
  {\doibase 10.1103/PhysRevB.102.081112} {\bibfield  {journal} {\bibinfo
  {journal} {Phys. Rev. B}\ }\textbf {\bibinfo {volume} {102}},\ \bibinfo
  {pages} {081112} (\bibinfo {year} {2020})},\ \Eprint
  {http://arxiv.org/abs/2001.09155} {arXiv:2001.09155 [cond-mat.str-el]}
  \BibitemShut {NoStop}%
\bibitem [{\citenamefont {Ray}\ and\ \citenamefont
  {Janssen}(2021)}]{Ray:2021moi}%
  \BibitemOpen
  \bibfield  {author} {\bibinfo {author} {\bibfnamefont {S.}~\bibnamefont
  {Ray}}\ and\ \bibinfo {author} {\bibfnamefont {L.}~\bibnamefont {Janssen}},\
  }\href {\doibase 10.1103/PhysRevB.104.045101} {\bibfield  {journal} {\bibinfo
   {journal} {Phys. Rev. B}\ }\textbf {\bibinfo {volume} {104}},\ \bibinfo
  {pages} {045101} (\bibinfo {year} {2021})},\ \Eprint
  {http://arxiv.org/abs/2104.12779} {arXiv:2104.12779 [cond-mat.str-el]}
  \BibitemShut {NoStop}%
\bibitem [{\citenamefont {Ray}(2022)}]{Ray:PhD}%
  \BibitemOpen
  \bibfield  {author} {\bibinfo {author} {\bibfnamefont {S.}~\bibnamefont
  {Ray}},\ }\emph {\bibinfo {title} {{Emergence and breakdown of quantum scale
  symmetry: from correlated condensed matter to physics beyond the standard
  mode (Ph.D. thesis)}}},\ \href
  {https://katalog.slub-dresden.de/id/0-1819524388} {Ph.D. thesis},\ \bibinfo
  {school} {TU Dresden} (\bibinfo {year} {2022})\BibitemShut {NoStop}%
\bibitem [{\citenamefont {Dey}\ and\ \citenamefont
  {Maciejko}(2022)}]{Dey:2022lkx}%
  \BibitemOpen
  \bibfield  {author} {\bibinfo {author} {\bibfnamefont {S.}~\bibnamefont
  {Dey}}\ and\ \bibinfo {author} {\bibfnamefont {J.}~\bibnamefont {Maciejko}},\
  }\href {\doibase 10.1103/PhysRevB.106.035140} {\bibfield  {journal} {\bibinfo
   {journal} {Phys. Rev. B}\ }\textbf {\bibinfo {volume} {106}},\ \bibinfo
  {pages} {035140} (\bibinfo {year} {2022})},\ \Eprint
  {http://arxiv.org/abs/2204.05319} {arXiv:2204.05319 [cond-mat.str-el]}
  \BibitemShut {NoStop}%
\bibitem [{\citenamefont {Gies}\ \emph {et~al.}(2024)\citenamefont {Gies},
  \citenamefont {Heinzel}, \citenamefont {Laufk\"otter},\ and\ \citenamefont
  {Picciau}}]{Gies:2023cnd}%
  \BibitemOpen
  \bibfield  {author} {\bibinfo {author} {\bibfnamefont {H.}~\bibnamefont
  {Gies}}, \bibinfo {author} {\bibfnamefont {P.}~\bibnamefont {Heinzel}},
  \bibinfo {author} {\bibfnamefont {J.}~\bibnamefont {Laufk\"otter}}, \ and\
  \bibinfo {author} {\bibfnamefont {M.}~\bibnamefont {Picciau}},\ }\href
  {\doibase 10.1103/PhysRevD.110.065001} {\bibfield  {journal} {\bibinfo
  {journal} {Phys. Rev. D}\ }\textbf {\bibinfo {volume} {110}},\ \bibinfo
  {pages} {065001} (\bibinfo {year} {2024})},\ \Eprint
  {http://arxiv.org/abs/2312.12058} {arXiv:2312.12058 [hep-th]} \BibitemShut
  {NoStop}%
\bibitem [{\citenamefont {Pais}\ and\ \citenamefont
  {Uhlenbeck}(1950)}]{Pais:1950za}%
  \BibitemOpen
  \bibfield  {author} {\bibinfo {author} {\bibfnamefont {A.}~\bibnamefont
  {Pais}}\ and\ \bibinfo {author} {\bibfnamefont {G.~E.}\ \bibnamefont
  {Uhlenbeck}},\ }\href {\doibase 10.1103/PhysRev.79.145} {\bibfield  {journal}
  {\bibinfo  {journal} {Phys. Rev.}\ }\textbf {\bibinfo {volume} {79}},\
  \bibinfo {pages} {145} (\bibinfo {year} {1950})}\BibitemShut {NoStop}%
\bibitem [{\citenamefont {Lee}\ and\ \citenamefont {Wick}(1970)}]{Lee:1970iw}%
  \BibitemOpen
  \bibfield  {author} {\bibinfo {author} {\bibfnamefont {T.~D.}\ \bibnamefont
  {Lee}}\ and\ \bibinfo {author} {\bibfnamefont {G.~C.}\ \bibnamefont {Wick}},\
  }\href {\doibase 10.1103/PhysRevD.2.1033} {\bibfield  {journal} {\bibinfo
  {journal} {Phys. Rev. D}\ }\textbf {\bibinfo {volume} {2}},\ \bibinfo {pages}
  {1033} (\bibinfo {year} {1970})}\BibitemShut {NoStop}%
\bibitem [{\citenamefont {Stelle}(1977)}]{Stelle:1976gc}%
  \BibitemOpen
  \bibfield  {author} {\bibinfo {author} {\bibfnamefont {K.~S.}\ \bibnamefont
  {Stelle}},\ }\href {\doibase 10.1103/PhysRevD.16.953} {\bibfield  {journal}
  {\bibinfo  {journal} {Phys. Rev. D}\ }\textbf {\bibinfo {volume} {16}},\
  \bibinfo {pages} {953} (\bibinfo {year} {1977})}\BibitemShut {NoStop}%
\bibitem [{\citenamefont {Grinstein}\ \emph {et~al.}(2008)\citenamefont
  {Grinstein}, \citenamefont {O'Connell},\ and\ \citenamefont
  {Wise}}]{Grinstein:2007mp}%
  \BibitemOpen
  \bibfield  {author} {\bibinfo {author} {\bibfnamefont {B.}~\bibnamefont
  {Grinstein}}, \bibinfo {author} {\bibfnamefont {D.}~\bibnamefont
  {O'Connell}}, \ and\ \bibinfo {author} {\bibfnamefont {M.~B.}\ \bibnamefont
  {Wise}},\ }\href {\doibase 10.1103/PhysRevD.77.025012} {\bibfield  {journal}
  {\bibinfo  {journal} {Phys. Rev. D}\ }\textbf {\bibinfo {volume} {77}},\
  \bibinfo {pages} {025012} (\bibinfo {year} {2008})},\ \Eprint
  {http://arxiv.org/abs/0704.1845} {arXiv:0704.1845 [hep-ph]} \BibitemShut
  {NoStop}%
\bibitem [{\citenamefont {Ostrogradsky}(1850)}]{Ostrogradsky:1850fid}%
  \BibitemOpen
  \bibfield  {author} {\bibinfo {author} {\bibfnamefont {M.}~\bibnamefont
  {Ostrogradsky}},\ }\href@noop {} {\bibfield  {journal} {\bibinfo  {journal}
  {Mem. Acad. St. Petersbourg}\ }\textbf {\bibinfo {volume} {6}},\ \bibinfo
  {pages} {385} (\bibinfo {year} {1850})}\BibitemShut {NoStop}%
\bibitem [{\citenamefont {Narnhofer}\ and\ \citenamefont
  {Thirring}(1978)}]{Narnhofer:1978sw}%
  \BibitemOpen
  \bibfield  {author} {\bibinfo {author} {\bibfnamefont {H.}~\bibnamefont
  {Narnhofer}}\ and\ \bibinfo {author} {\bibfnamefont {W.~E.}\ \bibnamefont
  {Thirring}},\ }\href {\doibase 10.1016/0370-2693(78)90898-5} {\bibfield
  {journal} {\bibinfo  {journal} {Phys. Lett. B}\ }\textbf {\bibinfo {volume}
  {76}},\ \bibinfo {pages} {428} (\bibinfo {year} {1978})}\BibitemShut
  {NoStop}%
\bibitem [{\citenamefont {Hawking}\ and\ \citenamefont
  {Hertog}(2002)}]{Hawking:2001yt}%
  \BibitemOpen
  \bibfield  {author} {\bibinfo {author} {\bibfnamefont {S.~W.}\ \bibnamefont
  {Hawking}}\ and\ \bibinfo {author} {\bibfnamefont {T.}~\bibnamefont
  {Hertog}},\ }\href {\doibase 10.1103/PhysRevD.65.103515} {\bibfield
  {journal} {\bibinfo  {journal} {Phys. Rev. D}\ }\textbf {\bibinfo {volume}
  {65}},\ \bibinfo {pages} {103515} (\bibinfo {year} {2002})},\ \Eprint
  {http://arxiv.org/abs/hep-th/0107088} {arXiv:hep-th/0107088} \BibitemShut
  {NoStop}%
\bibitem [{\citenamefont {Bender}\ and\ \citenamefont
  {Mannheim}(2008)}]{Bender:2007wu}%
  \BibitemOpen
  \bibfield  {author} {\bibinfo {author} {\bibfnamefont {C.~M.}\ \bibnamefont
  {Bender}}\ and\ \bibinfo {author} {\bibfnamefont {P.~D.}\ \bibnamefont
  {Mannheim}},\ }\href {\doibase 10.1103/PhysRevLett.100.110402} {\bibfield
  {journal} {\bibinfo  {journal} {Phys. Rev. Lett.}\ }\textbf {\bibinfo
  {volume} {100}},\ \bibinfo {pages} {110402} (\bibinfo {year} {2008})},\
  \Eprint {http://arxiv.org/abs/0706.0207} {arXiv:0706.0207 [hep-th]}
  \BibitemShut {NoStop}%
\bibitem [{\citenamefont {Garriga}\ and\ \citenamefont
  {Vilenkin}(2013)}]{Garriga:2012pk}%
  \BibitemOpen
  \bibfield  {author} {\bibinfo {author} {\bibfnamefont {J.}~\bibnamefont
  {Garriga}}\ and\ \bibinfo {author} {\bibfnamefont {A.}~\bibnamefont
  {Vilenkin}},\ }\href {\doibase 10.1088/1475-7516/2013/01/036} {\bibfield
  {journal} {\bibinfo  {journal} {JCAP}\ }\textbf {\bibinfo {volume} {01}},\
  \bibinfo {pages} {036} (\bibinfo {year} {2013})},\ \Eprint
  {http://arxiv.org/abs/1202.1239} {arXiv:1202.1239 [hep-th]} \BibitemShut
  {NoStop}%
\bibitem [{\citenamefont {Salvio}\ and\ \citenamefont
  {Strumia}(2014)}]{Salvio:2014soa}%
  \BibitemOpen
  \bibfield  {author} {\bibinfo {author} {\bibfnamefont {A.}~\bibnamefont
  {Salvio}}\ and\ \bibinfo {author} {\bibfnamefont {A.}~\bibnamefont
  {Strumia}},\ }\href {\doibase 10.1007/JHEP06(2014)080} {\bibfield  {journal}
  {\bibinfo  {journal} {JHEP}\ }\textbf {\bibinfo {volume} {06}},\ \bibinfo
  {pages} {080} (\bibinfo {year} {2014})},\ \Eprint
  {http://arxiv.org/abs/1403.4226} {arXiv:1403.4226 [hep-ph]} \BibitemShut
  {NoStop}%
\bibitem [{\citenamefont {Smilga}(2017)}]{Smilga:2017arl}%
  \BibitemOpen
  \bibfield  {author} {\bibinfo {author} {\bibfnamefont {A.}~\bibnamefont
  {Smilga}},\ }\href {\doibase 10.1142/S0217751X17300253} {\bibfield  {journal}
  {\bibinfo  {journal} {Int. J. Mod. Phys. A}\ }\textbf {\bibinfo {volume}
  {32}},\ \bibinfo {pages} {1730025} (\bibinfo {year} {2017})},\ \Eprint
  {http://arxiv.org/abs/1710.11538} {arXiv:1710.11538 [hep-th]} \BibitemShut
  {NoStop}%
\bibitem [{\citenamefont {Becker}\ \emph {et~al.}(2017)\citenamefont {Becker},
  \citenamefont {Ripken},\ and\ \citenamefont {Saueressig}}]{Becker:2017tcx}%
  \BibitemOpen
  \bibfield  {author} {\bibinfo {author} {\bibfnamefont {D.}~\bibnamefont
  {Becker}}, \bibinfo {author} {\bibfnamefont {C.}~\bibnamefont {Ripken}}, \
  and\ \bibinfo {author} {\bibfnamefont {F.}~\bibnamefont {Saueressig}},\
  }\href {\doibase 10.1007/JHEP12(2017)121} {\bibfield  {journal} {\bibinfo
  {journal} {JHEP}\ }\textbf {\bibinfo {volume} {12}},\ \bibinfo {pages} {121}
  (\bibinfo {year} {2017})},\ \Eprint {http://arxiv.org/abs/1709.09098}
  {arXiv:1709.09098 [hep-th]} \BibitemShut {NoStop}%
\bibitem [{\citenamefont {Anselmi}(2018)}]{Anselmi:2018kgz}%
  \BibitemOpen
  \bibfield  {author} {\bibinfo {author} {\bibfnamefont {D.}~\bibnamefont
  {Anselmi}},\ }\href {\doibase 10.1007/JHEP02(2018)141} {\bibfield  {journal}
  {\bibinfo  {journal} {JHEP}\ }\textbf {\bibinfo {volume} {02}},\ \bibinfo
  {pages} {141} (\bibinfo {year} {2018})},\ \Eprint
  {http://arxiv.org/abs/1801.00915} {arXiv:1801.00915 [hep-th]} \BibitemShut
  {NoStop}%
\bibitem [{\citenamefont {Gross}\ \emph {et~al.}(2021)\citenamefont {Gross},
  \citenamefont {Strumia}, \citenamefont {Teresi},\ and\ \citenamefont
  {Zirilli}}]{Gross:2020tph}%
  \BibitemOpen
  \bibfield  {author} {\bibinfo {author} {\bibfnamefont {C.}~\bibnamefont
  {Gross}}, \bibinfo {author} {\bibfnamefont {A.}~\bibnamefont {Strumia}},
  \bibinfo {author} {\bibfnamefont {D.}~\bibnamefont {Teresi}}, \ and\ \bibinfo
  {author} {\bibfnamefont {M.}~\bibnamefont {Zirilli}},\ }\href {\doibase
  10.1103/PhysRevD.103.115025} {\bibfield  {journal} {\bibinfo  {journal}
  {Phys. Rev. D}\ }\textbf {\bibinfo {volume} {103}},\ \bibinfo {pages}
  {115025} (\bibinfo {year} {2021})},\ \Eprint
  {http://arxiv.org/abs/2007.05541} {arXiv:2007.05541 [hep-th]} \BibitemShut
  {NoStop}%
\bibitem [{\citenamefont {Donoghue}\ and\ \citenamefont
  {Menezes}(2021)}]{Donoghue:2021eto}%
  \BibitemOpen
  \bibfield  {author} {\bibinfo {author} {\bibfnamefont {J.~F.}\ \bibnamefont
  {Donoghue}}\ and\ \bibinfo {author} {\bibfnamefont {G.}~\bibnamefont
  {Menezes}},\ }\href {\doibase 10.1103/PhysRevD.104.045010} {\bibfield
  {journal} {\bibinfo  {journal} {Phys. Rev. D}\ }\textbf {\bibinfo {volume}
  {104}},\ \bibinfo {pages} {045010} (\bibinfo {year} {2021})},\ \Eprint
  {http://arxiv.org/abs/2105.00898} {arXiv:2105.00898 [hep-th]} \BibitemShut
  {NoStop}%
\bibitem [{\citenamefont {Platania}(2019)}]{Platania:2019qvo}%
  \BibitemOpen
  \bibfield  {author} {\bibinfo {author} {\bibfnamefont {A.}~\bibnamefont
  {Platania}},\ }\href {\doibase 10.3390/universe5080189} {\bibfield  {journal}
  {\bibinfo  {journal} {Universe}\ }\textbf {\bibinfo {volume} {5}},\ \bibinfo
  {pages} {189} (\bibinfo {year} {2019})},\ \Eprint
  {http://arxiv.org/abs/1908.03897} {arXiv:1908.03897 [gr-qc]} \BibitemShut
  {NoStop}%
\bibitem [{\citenamefont {Deffayet}\ \emph {et~al.}(2023)\citenamefont
  {Deffayet}, \citenamefont {Held}, \citenamefont {Mukohyama},\ and\
  \citenamefont {Vikman}}]{Deffayet:2023wdg}%
  \BibitemOpen
  \bibfield  {author} {\bibinfo {author} {\bibfnamefont {C.}~\bibnamefont
  {Deffayet}}, \bibinfo {author} {\bibfnamefont {A.}~\bibnamefont {Held}},
  \bibinfo {author} {\bibfnamefont {S.}~\bibnamefont {Mukohyama}}, \ and\
  \bibinfo {author} {\bibfnamefont {A.}~\bibnamefont {Vikman}},\ }\href
  {\doibase 10.1088/1475-7516/2023/11/031} {\bibfield  {journal} {\bibinfo
  {journal} {JCAP}\ }\textbf {\bibinfo {volume} {11}},\ \bibinfo {pages} {031}
  (\bibinfo {year} {2023})},\ \Eprint {http://arxiv.org/abs/2305.09631}
  {arXiv:2305.09631 [gr-qc]} \BibitemShut {NoStop}%
\bibitem [{\citenamefont {Coleman}\ and\ \citenamefont
  {Weinberg}(1973)}]{Coleman:1973jx}%
  \BibitemOpen
  \bibfield  {author} {\bibinfo {author} {\bibfnamefont {S.~R.}\ \bibnamefont
  {Coleman}}\ and\ \bibinfo {author} {\bibfnamefont {E.~J.}\ \bibnamefont
  {Weinberg}},\ }\href {\doibase 10.1103/PhysRevD.7.1888} {\bibfield  {journal}
  {\bibinfo  {journal} {Phys. Rev.}\ }\textbf {\bibinfo {volume} {D7}},\
  \bibinfo {pages} {1888} (\bibinfo {year} {1973})}\BibitemShut {NoStop}%
\bibitem [{\citenamefont {Semenoff}(1984)}]{Semenoff:1984dq}%
  \BibitemOpen
  \bibfield  {author} {\bibinfo {author} {\bibfnamefont {G.~W.}\ \bibnamefont
  {Semenoff}},\ }\href {\doibase 10.1103/PhysRevLett.53.2449} {\bibfield
  {journal} {\bibinfo  {journal} {Phys. Rev. Lett.}\ }\textbf {\bibinfo
  {volume} {53}},\ \bibinfo {pages} {2449} (\bibinfo {year}
  {1984})}\BibitemShut {NoStop}%
\bibitem [{\citenamefont {Appelquist}\ \emph {et~al.}(1986)\citenamefont
  {Appelquist}, \citenamefont {Bowick}, \citenamefont {Karabali},\ and\
  \citenamefont {Wijewardhana}}]{Appelquist:1986fd}%
  \BibitemOpen
  \bibfield  {author} {\bibinfo {author} {\bibfnamefont {T.~W.}\ \bibnamefont
  {Appelquist}}, \bibinfo {author} {\bibfnamefont {M.~J.}\ \bibnamefont
  {Bowick}}, \bibinfo {author} {\bibfnamefont {D.}~\bibnamefont {Karabali}}, \
  and\ \bibinfo {author} {\bibfnamefont {L.~C.~R.}\ \bibnamefont
  {Wijewardhana}},\ }\href {\doibase 10.1103/PhysRevD.33.3704} {\bibfield
  {journal} {\bibinfo  {journal} {Phys. Rev. D}\ }\textbf {\bibinfo {volume}
  {33}},\ \bibinfo {pages} {3704} (\bibinfo {year} {1986})}\BibitemShut
  {NoStop}%
\bibitem [{\citenamefont {Haldane}(1988)}]{Haldane:1988zza}%
  \BibitemOpen
  \bibfield  {author} {\bibinfo {author} {\bibfnamefont {F.~D.~M.}\
  \bibnamefont {Haldane}},\ }\href {\doibase 10.1103/PhysRevLett.61.2015}
  {\bibfield  {journal} {\bibinfo  {journal} {Phys. Rev. Lett.}\ }\textbf
  {\bibinfo {volume} {61}},\ \bibinfo {pages} {2015} (\bibinfo {year}
  {1988})}\BibitemShut {NoStop}%
\bibitem [{\citenamefont {Kane}\ and\ \citenamefont
  {Mele}(2005)}]{Kane:2004bvs}%
  \BibitemOpen
  \bibfield  {author} {\bibinfo {author} {\bibfnamefont {C.~L.}\ \bibnamefont
  {Kane}}\ and\ \bibinfo {author} {\bibfnamefont {E.~J.}\ \bibnamefont
  {Mele}},\ }\href {\doibase 10.1103/PhysRevLett.95.226801} {\bibfield
  {journal} {\bibinfo  {journal} {Phys. Rev. Lett.}\ }\textbf {\bibinfo
  {volume} {95}},\ \bibinfo {pages} {226801} (\bibinfo {year} {2005})},\
  \Eprint {http://arxiv.org/abs/cond-mat/0411737} {arXiv:cond-mat/0411737}
  \BibitemShut {NoStop}%
\bibitem [{\citenamefont {Schrödinger}(1932)}]{Schroedinger:1932a}%
  \BibitemOpen
  \bibfield  {author} {\bibinfo {author} {\bibfnamefont {E.}~\bibnamefont
  {Schrödinger}},\ }\href {\doibase 10.34663/9783945561317-15} {\bibfield
  {journal} {\bibinfo  {journal} {Sitz.ber. Preuss. Akad. Wiss. (Berlin),
  Phys.-math. Kl.}\ ,\ \bibinfo {pages} {105}} (\bibinfo {year}
  {1932})}\BibitemShut {NoStop}%
\bibitem [{\citenamefont {Bargmann}(1932)}]{Bargmann:1932a}%
  \BibitemOpen
  \bibfield  {author} {\bibinfo {author} {\bibfnamefont {V.}~\bibnamefont
  {Bargmann}},\ }\href@noop {} {\bibfield  {journal} {\bibinfo  {journal}
  {Sitz.ber. Preuss. Akad. Wiss. (Berlin), Phys.-math. Kl.}\ ,\ \bibinfo
  {pages} {346}} (\bibinfo {year} {1932})}\BibitemShut {NoStop}%
\bibitem [{\citenamefont {Weldon}(2001)}]{Weldon:2000fr}%
  \BibitemOpen
  \bibfield  {author} {\bibinfo {author} {\bibfnamefont {H.~A.}\ \bibnamefont
  {Weldon}},\ }\href {\doibase 10.1103/PhysRevD.63.104010} {\bibfield
  {journal} {\bibinfo  {journal} {Phys. Rev. D}\ }\textbf {\bibinfo {volume}
  {63}},\ \bibinfo {pages} {104010} (\bibinfo {year} {2001})},\ \Eprint
  {http://arxiv.org/abs/gr-qc/0009086} {arXiv:gr-qc/0009086} \BibitemShut
  {NoStop}%
\bibitem [{\citenamefont {Gies}\ and\ \citenamefont
  {Lippoldt}(2014)}]{Gies:2013noa}%
  \BibitemOpen
  \bibfield  {author} {\bibinfo {author} {\bibfnamefont {H.}~\bibnamefont
  {Gies}}\ and\ \bibinfo {author} {\bibfnamefont {S.}~\bibnamefont
  {Lippoldt}},\ }\href {\doibase 10.1103/PhysRevD.89.064040} {\bibfield
  {journal} {\bibinfo  {journal} {Phys. Rev. D}\ }\textbf {\bibinfo {volume}
  {89}},\ \bibinfo {pages} {064040} (\bibinfo {year} {2014})},\ \Eprint
  {http://arxiv.org/abs/1310.2509} {arXiv:1310.2509 [hep-th]} \BibitemShut
  {NoStop}%
\bibitem [{\citenamefont {Schiffhorst}()}]{Schiffhorst:2024}%
  \BibitemOpen
  \bibfield  {author} {\bibinfo {author} {\bibfnamefont {L.}~\bibnamefont
  {Schiffhorst}},\ }\href@noop {} {\enquote {\bibinfo {title} {{Bachelor
  thesis}},}\ }\bibinfo {note} {{Jena, (2024)}}\BibitemShut {NoStop}%
\bibitem [{\citenamefont {Gross}\ and\ \citenamefont
  {Neveu}(1974)}]{Gross:1974jv}%
  \BibitemOpen
  \bibfield  {author} {\bibinfo {author} {\bibfnamefont {D.~J.}\ \bibnamefont
  {Gross}}\ and\ \bibinfo {author} {\bibfnamefont {A.}~\bibnamefont {Neveu}},\
  }\href {\doibase 10.1103/PhysRevD.10.3235} {\bibfield  {journal} {\bibinfo
  {journal} {Phys. Rev. D}\ }\textbf {\bibinfo {volume} {10}},\ \bibinfo
  {pages} {3235} (\bibinfo {year} {1974})}\BibitemShut {NoStop}%
\bibitem [{\citenamefont {Hofling}\ \emph {et~al.}(2002)\citenamefont
  {Hofling}, \citenamefont {Nowak},\ and\ \citenamefont
  {Wetterich}}]{Hofling:2002hj}%
  \BibitemOpen
  \bibfield  {author} {\bibinfo {author} {\bibfnamefont {F.}~\bibnamefont
  {Hofling}}, \bibinfo {author} {\bibfnamefont {C.}~\bibnamefont {Nowak}}, \
  and\ \bibinfo {author} {\bibfnamefont {C.}~\bibnamefont {Wetterich}},\ }\href
  {\doibase 10.1103/PhysRevB.66.205111} {\bibfield  {journal} {\bibinfo
  {journal} {Phys. Rev.}\ }\textbf {\bibinfo {volume} {B66}},\ \bibinfo {pages}
  {205111} (\bibinfo {year} {2002})},\ \Eprint
  {http://arxiv.org/abs/cond-mat/0203588} {arXiv:cond-mat/0203588 [cond-mat]}
  \BibitemShut {NoStop}%
\bibitem [{\citenamefont {Nambu}\ and\ \citenamefont
  {Jona-Lasinio}(1961)}]{Nambu:1961tp}%
  \BibitemOpen
  \bibfield  {author} {\bibinfo {author} {\bibfnamefont {Y.}~\bibnamefont
  {Nambu}}\ and\ \bibinfo {author} {\bibfnamefont {G.}~\bibnamefont
  {Jona-Lasinio}},\ }\href {\doibase 10.1103/PhysRev.122.345} {\bibfield
  {journal} {\bibinfo  {journal} {Phys. Rev.}\ }\textbf {\bibinfo {volume}
  {122}},\ \bibinfo {pages} {345} (\bibinfo {year} {1961})}\BibitemShut
  {NoStop}%
\bibitem [{\citenamefont {Wetterich}(1993)}]{Wetterich:1992yh}%
  \BibitemOpen
  \bibfield  {author} {\bibinfo {author} {\bibfnamefont {C.}~\bibnamefont
  {Wetterich}},\ }\href {\doibase 10.1016/0370-2693(93)90726-X} {\bibfield
  {journal} {\bibinfo  {journal} {Phys. Lett.}\ }\textbf {\bibinfo {volume}
  {B301}},\ \bibinfo {pages} {90} (\bibinfo {year} {1993})}\BibitemShut
  {NoStop}%
\bibitem [{\citenamefont {Gies}\ and\ \citenamefont
  {Wetterich}(2002)}]{Gies:2001nw}%
  \BibitemOpen
  \bibfield  {author} {\bibinfo {author} {\bibfnamefont {H.}~\bibnamefont
  {Gies}}\ and\ \bibinfo {author} {\bibfnamefont {C.}~\bibnamefont
  {Wetterich}},\ }\href {\doibase 10.1103/PhysRevD.65.065001} {\bibfield
  {journal} {\bibinfo  {journal} {Phys. Rev.}\ }\textbf {\bibinfo {volume}
  {D65}},\ \bibinfo {pages} {065001} (\bibinfo {year} {2002})},\ \Eprint
  {http://arxiv.org/abs/hep-th/0107221} {arXiv:hep-th/0107221 [hep-th]}
  \BibitemShut {NoStop}%
\bibitem [{\citenamefont {Braun}(2012)}]{Braun:2011pp}%
  \BibitemOpen
  \bibfield  {author} {\bibinfo {author} {\bibfnamefont {J.}~\bibnamefont
  {Braun}},\ }\href {\doibase 10.1088/0954-3899/39/3/033001} {\bibfield
  {journal} {\bibinfo  {journal} {J. Phys.}\ }\textbf {\bibinfo {volume}
  {G39}},\ \bibinfo {pages} {033001} (\bibinfo {year} {2012})},\ \Eprint
  {http://arxiv.org/abs/1108.4449} {arXiv:1108.4449 [hep-ph]} \BibitemShut
  {NoStop}%
\bibitem [{\citenamefont {Gehring}\ \emph {et~al.}(2015)\citenamefont
  {Gehring}, \citenamefont {Gies},\ and\ \citenamefont
  {Janssen}}]{Gehring:2015vja}%
  \BibitemOpen
  \bibfield  {author} {\bibinfo {author} {\bibfnamefont {F.}~\bibnamefont
  {Gehring}}, \bibinfo {author} {\bibfnamefont {H.}~\bibnamefont {Gies}}, \
  and\ \bibinfo {author} {\bibfnamefont {L.}~\bibnamefont {Janssen}},\ }\href
  {\doibase 10.1103/PhysRevD.92.085046} {\bibfield  {journal} {\bibinfo
  {journal} {Phys. Rev.}\ }\textbf {\bibinfo {volume} {D92}},\ \bibinfo {pages}
  {085046} (\bibinfo {year} {2015})},\ \Eprint
  {http://arxiv.org/abs/1506.07570} {arXiv:1506.07570 [hep-th]} \BibitemShut
  {NoStop}%
\bibitem [{\citenamefont {Braun}\ \emph {et~al.}(2011)\citenamefont {Braun},
  \citenamefont {Gies},\ and\ \citenamefont {Scherer}}]{Braun:2010tt}%
  \BibitemOpen
  \bibfield  {author} {\bibinfo {author} {\bibfnamefont {J.}~\bibnamefont
  {Braun}}, \bibinfo {author} {\bibfnamefont {H.}~\bibnamefont {Gies}}, \ and\
  \bibinfo {author} {\bibfnamefont {D.~D.}\ \bibnamefont {Scherer}},\ }\href
  {\doibase 10.1103/PhysRevD.83.085012} {\bibfield  {journal} {\bibinfo
  {journal} {Phys. Rev.}\ }\textbf {\bibinfo {volume} {D83}},\ \bibinfo {pages}
  {085012} (\bibinfo {year} {2011})},\ \Eprint {http://arxiv.org/abs/1011.1456}
  {arXiv:1011.1456 [hep-th]} \BibitemShut {NoStop}%
\bibitem [{\citenamefont {Mesterhazy}\ \emph {et~al.}(2012)\citenamefont
  {Mesterhazy}, \citenamefont {Berges},\ and\ \citenamefont {von
  Smekal}}]{Mesterhazy:2012ei}%
  \BibitemOpen
  \bibfield  {author} {\bibinfo {author} {\bibfnamefont {D.}~\bibnamefont
  {Mesterhazy}}, \bibinfo {author} {\bibfnamefont {J.}~\bibnamefont {Berges}},
  \ and\ \bibinfo {author} {\bibfnamefont {L.}~\bibnamefont {von Smekal}},\
  }\href {\doibase 10.1103/PhysRevB.86.245431} {\bibfield  {journal} {\bibinfo
  {journal} {Phys. Rev.}\ }\textbf {\bibinfo {volume} {B86}},\ \bibinfo {pages}
  {245431} (\bibinfo {year} {2012})},\ \Eprint {http://arxiv.org/abs/1207.4054}
  {arXiv:1207.4054 [cond-mat.str-el]} \BibitemShut {NoStop}%
\bibitem [{\citenamefont {Jakov\'{a}c}\ \emph {et~al.}(2015)\citenamefont
  {Jakov\'{a}c}, \citenamefont {Patk\'{o}s},\ and\ \citenamefont
  {P\'{o}sfay}}]{Jakovac:2014lqa}%
  \BibitemOpen
  \bibfield  {author} {\bibinfo {author} {\bibfnamefont {A.}~\bibnamefont
  {Jakov\'{a}c}}, \bibinfo {author} {\bibfnamefont {A.}~\bibnamefont
  {Patk\'{o}s}}, \ and\ \bibinfo {author} {\bibfnamefont {P.}~\bibnamefont
  {P\'{o}sfay}},\ }\href {\doibase 10.1140/epjc/s10052-014-3228-1} {\bibfield
  {journal} {\bibinfo  {journal} {Eur. Phys. J.}\ }\textbf {\bibinfo {volume}
  {C75}},\ \bibinfo {pages} {2} (\bibinfo {year} {2015})},\ \Eprint
  {http://arxiv.org/abs/1406.3195} {arXiv:1406.3195 [hep-th]} \BibitemShut
  {NoStop}%
\bibitem [{\citenamefont {Janssen}\ and\ \citenamefont
  {Herbut}(2014)}]{Janssen:2014gea}%
  \BibitemOpen
  \bibfield  {author} {\bibinfo {author} {\bibfnamefont {L.}~\bibnamefont
  {Janssen}}\ and\ \bibinfo {author} {\bibfnamefont {I.~F.}\ \bibnamefont
  {Herbut}},\ }\href {\doibase 10.1103/PhysRevB.89.205403} {\bibfield
  {journal} {\bibinfo  {journal} {Phys. Rev.}\ }\textbf {\bibinfo {volume}
  {B89}},\ \bibinfo {pages} {205403} (\bibinfo {year} {2014})},\ \Eprint
  {http://arxiv.org/abs/1402.6277} {arXiv:1402.6277 [cond-mat.str-el]}
  \BibitemShut {NoStop}%
\bibitem [{\citenamefont {Vacca}\ and\ \citenamefont
  {Zambelli}(2015)}]{Vacca:2015nta}%
  \BibitemOpen
  \bibfield  {author} {\bibinfo {author} {\bibfnamefont {G.~P.}\ \bibnamefont
  {Vacca}}\ and\ \bibinfo {author} {\bibfnamefont {L.}~\bibnamefont
  {Zambelli}},\ }\href {\doibase 10.1103/PhysRevD.91.125003} {\bibfield
  {journal} {\bibinfo  {journal} {Phys. Rev.}\ }\textbf {\bibinfo {volume}
  {D91}},\ \bibinfo {pages} {125003} (\bibinfo {year} {2015})},\ \Eprint
  {http://arxiv.org/abs/1503.09136} {arXiv:1503.09136 [hep-th]} \BibitemShut
  {NoStop}%
\bibitem [{\citenamefont {Classen}\ \emph {et~al.}(2016)\citenamefont
  {Classen}, \citenamefont {Herbut}, \citenamefont {Janssen},\ and\
  \citenamefont {Scherer}}]{Classen:2015mar}%
  \BibitemOpen
  \bibfield  {author} {\bibinfo {author} {\bibfnamefont {L.}~\bibnamefont
  {Classen}}, \bibinfo {author} {\bibfnamefont {I.~F.}\ \bibnamefont {Herbut}},
  \bibinfo {author} {\bibfnamefont {L.}~\bibnamefont {Janssen}}, \ and\
  \bibinfo {author} {\bibfnamefont {M.~M.}\ \bibnamefont {Scherer}},\ }\href
  {\doibase 10.1103/PhysRevB.93.125119} {\bibfield  {journal} {\bibinfo
  {journal} {Phys. Rev.}\ }\textbf {\bibinfo {volume} {B93}},\ \bibinfo {pages}
  {125119} (\bibinfo {year} {2016})},\ \Eprint
  {http://arxiv.org/abs/1510.09003} {arXiv:1510.09003 [cond-mat.str-el]}
  \BibitemShut {NoStop}%
\bibitem [{\citenamefont {Knorr}(2016)}]{Knorr:2016sfs}%
  \BibitemOpen
  \bibfield  {author} {\bibinfo {author} {\bibfnamefont {B.}~\bibnamefont
  {Knorr}},\ }\href {\doibase 10.1103/PhysRevB.94.245102} {\bibfield  {journal}
  {\bibinfo  {journal} {Phys. Rev.}\ }\textbf {\bibinfo {volume} {B94}},\
  \bibinfo {pages} {245102} (\bibinfo {year} {2016})},\ \Eprint
  {http://arxiv.org/abs/1609.03824} {arXiv:1609.03824 [cond-mat.str-el]}
  \BibitemShut {NoStop}%
\bibitem [{\citenamefont {Cresswell-Hogg}\ and\ \citenamefont
  {Litim}(2023{\natexlab{a}})}]{Cresswell-Hogg:2022lgg}%
  \BibitemOpen
  \bibfield  {author} {\bibinfo {author} {\bibfnamefont {C.}~\bibnamefont
  {Cresswell-Hogg}}\ and\ \bibinfo {author} {\bibfnamefont {D.~F.}\
  \bibnamefont {Litim}},\ }\href {\doibase 10.1103/PhysRevLett.130.201602}
  {\bibfield  {journal} {\bibinfo  {journal} {Phys. Rev. Lett.}\ }\textbf
  {\bibinfo {volume} {130}},\ \bibinfo {pages} {201602} (\bibinfo {year}
  {2023}{\natexlab{a}})},\ \Eprint {http://arxiv.org/abs/2207.10115}
  {arXiv:2207.10115 [hep-th]} \BibitemShut {NoStop}%
\bibitem [{\citenamefont {Cresswell-Hogg}\ and\ \citenamefont
  {Litim}(2023{\natexlab{b}})}]{Cresswell-Hogg:2022lez}%
  \BibitemOpen
  \bibfield  {author} {\bibinfo {author} {\bibfnamefont {C.}~\bibnamefont
  {Cresswell-Hogg}}\ and\ \bibinfo {author} {\bibfnamefont {D.~F.}\
  \bibnamefont {Litim}},\ }\href {\doibase 10.1103/PhysRevD.107.L101701}
  {\bibfield  {journal} {\bibinfo  {journal} {Phys. Rev. D}\ }\textbf {\bibinfo
  {volume} {107}},\ \bibinfo {pages} {L101701} (\bibinfo {year}
  {2023}{\natexlab{b}})},\ \Eprint {http://arxiv.org/abs/2212.06815}
  {arXiv:2212.06815 [hep-th]} \BibitemShut {NoStop}%
\bibitem [{\citenamefont {Cresswell-Hogg}\ and\ \citenamefont
  {Litim}(2024)}]{Cresswell-Hogg:2023hdg}%
  \BibitemOpen
  \bibfield  {author} {\bibinfo {author} {\bibfnamefont {C.}~\bibnamefont
  {Cresswell-Hogg}}\ and\ \bibinfo {author} {\bibfnamefont {D.~F.}\
  \bibnamefont {Litim}},\ }\href {\doibase 10.1007/JHEP07(2024)066} {\bibfield
  {journal} {\bibinfo  {journal} {JHEP}\ }\textbf {\bibinfo {volume} {07}},\
  \bibinfo {pages} {066} (\bibinfo {year} {2024})},\ \Eprint
  {http://arxiv.org/abs/2311.16246} {arXiv:2311.16246 [hep-th]} \BibitemShut
  {NoStop}%
\bibitem [{\citenamefont {Heinzel}()}]{Heinzel:2023}%
  \BibitemOpen
  \bibfield  {author} {\bibinfo {author} {\bibfnamefont {P.}~\bibnamefont
  {Heinzel}},\ }\href@noop {} {\enquote {\bibinfo {title} {{Master thesis}},}\
  }\bibinfo {note} {{Jena, (2023)}}\BibitemShut {NoStop}%
\bibitem [{\citenamefont {Schwinger}(1951)}]{Schwinger:1951nm}%
  \BibitemOpen
  \bibfield  {author} {\bibinfo {author} {\bibfnamefont {J.~S.}\ \bibnamefont
  {Schwinger}},\ }\href {\doibase 10.1103/PhysRev.82.664} {\bibfield  {journal}
  {\bibinfo  {journal} {Phys. Rev.}\ }\textbf {\bibinfo {volume} {82}},\
  \bibinfo {pages} {664} (\bibinfo {year} {1951})}\BibitemShut {NoStop}%
\bibitem [{\citenamefont {Holland}(2005)}]{Holland:2004sd}%
  \BibitemOpen
  \bibfield  {author} {\bibinfo {author} {\bibfnamefont {K.}~\bibnamefont
  {Holland}},\ }\bibfield  {booktitle} {\emph {\bibinfo {booktitle} {{Lattice
  field theory. Proceedings, 22nd International Symposium, Lattice 2004,
  Batavia, USA, June 21-26, 2004}}},\ }\href {\doibase
  10.1016/j.nuclphysbps.2004.11.293} {\bibfield  {journal} {\bibinfo  {journal}
  {Nucl. Phys. Proc. Suppl.}\ }\textbf {\bibinfo {volume} {140}},\ \bibinfo
  {pages} {155} (\bibinfo {year} {2005})},\ \Eprint
  {http://arxiv.org/abs/hep-lat/0409112} {arXiv:hep-lat/0409112 [hep-lat]}
  \BibitemShut {NoStop}%
\bibitem [{\citenamefont {Gies}\ \emph {et~al.}(2014)\citenamefont {Gies},
  \citenamefont {Gneiting},\ and\ \citenamefont {Sondenheimer}}]{Gies:2013fua}%
  \BibitemOpen
  \bibfield  {author} {\bibinfo {author} {\bibfnamefont {H.}~\bibnamefont
  {Gies}}, \bibinfo {author} {\bibfnamefont {C.}~\bibnamefont {Gneiting}}, \
  and\ \bibinfo {author} {\bibfnamefont {R.}~\bibnamefont {Sondenheimer}},\
  }\href {\doibase 10.1103/PhysRevD.89.045012} {\bibfield  {journal} {\bibinfo
  {journal} {Phys. Rev.}\ }\textbf {\bibinfo {volume} {D89}},\ \bibinfo {pages}
  {045012} (\bibinfo {year} {2014})},\ \Eprint {http://arxiv.org/abs/1308.5075}
  {arXiv:1308.5075 [hep-ph]} \BibitemShut {NoStop}%
\bibitem [{\citenamefont {Gies}\ and\ \citenamefont
  {Sondenheimer}(2015)}]{Gies:2014xha}%
  \BibitemOpen
  \bibfield  {author} {\bibinfo {author} {\bibfnamefont {H.}~\bibnamefont
  {Gies}}\ and\ \bibinfo {author} {\bibfnamefont {R.}~\bibnamefont
  {Sondenheimer}},\ }\href {\doibase 10.1140/epjc/s10052-015-3284-1} {\bibfield
   {journal} {\bibinfo  {journal} {Eur. Phys. J.}\ }\textbf {\bibinfo {volume}
  {C75}},\ \bibinfo {pages} {68} (\bibinfo {year} {2015})},\ \Eprint
  {http://arxiv.org/abs/1407.8124} {arXiv:1407.8124 [hep-ph]} \BibitemShut
  {NoStop}%
\bibitem [{\citenamefont {Woodard}(2015)}]{Woodard:2015zca}%
  \BibitemOpen
  \bibfield  {author} {\bibinfo {author} {\bibfnamefont {R.~P.}\ \bibnamefont
  {Woodard}},\ }\href {\doibase 10.4249/scholarpedia.32243} {\bibfield
  {journal} {\bibinfo  {journal} {Scholarpedia}\ }\textbf {\bibinfo {volume}
  {10}},\ \bibinfo {pages} {32243} (\bibinfo {year} {2015})},\ \Eprint
  {http://arxiv.org/abs/1506.02210} {arXiv:1506.02210 [hep-th]} \BibitemShut
  {NoStop}%
\bibitem [{\citenamefont {Caldwell}(2002)}]{Caldwell:1999ew}%
  \BibitemOpen
  \bibfield  {author} {\bibinfo {author} {\bibfnamefont {R.~R.}\ \bibnamefont
  {Caldwell}},\ }\href {\doibase 10.1016/S0370-2693(02)02589-3} {\bibfield
  {journal} {\bibinfo  {journal} {Phys. Lett. B}\ }\textbf {\bibinfo {volume}
  {545}},\ \bibinfo {pages} {23} (\bibinfo {year} {2002})},\ \Eprint
  {http://arxiv.org/abs/astro-ph/9908168} {arXiv:astro-ph/9908168} \BibitemShut
  {NoStop}%
\bibitem [{\citenamefont {Cline}\ \emph {et~al.}(2004)\citenamefont {Cline},
  \citenamefont {Jeon},\ and\ \citenamefont {Moore}}]{Cline:2003gs}%
  \BibitemOpen
  \bibfield  {author} {\bibinfo {author} {\bibfnamefont {J.~M.}\ \bibnamefont
  {Cline}}, \bibinfo {author} {\bibfnamefont {S.}~\bibnamefont {Jeon}}, \ and\
  \bibinfo {author} {\bibfnamefont {G.~D.}\ \bibnamefont {Moore}},\ }\href
  {\doibase 10.1103/PhysRevD.70.043543} {\bibfield  {journal} {\bibinfo
  {journal} {Phys. Rev. D}\ }\textbf {\bibinfo {volume} {70}},\ \bibinfo
  {pages} {043543} (\bibinfo {year} {2004})},\ \Eprint
  {http://arxiv.org/abs/hep-ph/0311312} {arXiv:hep-ph/0311312} \BibitemShut
  {NoStop}%
\bibitem [{\citenamefont {Cline}\ \emph {et~al.}(2024)\citenamefont {Cline},
  \citenamefont {Puel},\ and\ \citenamefont {Toma}}]{Cline:2023hfw}%
  \BibitemOpen
  \bibfield  {author} {\bibinfo {author} {\bibfnamefont {J.~M.}\ \bibnamefont
  {Cline}}, \bibinfo {author} {\bibfnamefont {M.}~\bibnamefont {Puel}}, \ and\
  \bibinfo {author} {\bibfnamefont {T.}~\bibnamefont {Toma}},\ }\href {\doibase
  10.1016/j.physletb.2023.138377} {\bibfield  {journal} {\bibinfo  {journal}
  {Phys. Lett. B}\ }\textbf {\bibinfo {volume} {848}},\ \bibinfo {pages}
  {138377} (\bibinfo {year} {2024})},\ \Eprint
  {http://arxiv.org/abs/2308.01333} {arXiv:2308.01333 [hep-ph]} \BibitemShut
  {NoStop}%
\bibitem [{\citenamefont {Cline}(2024)}]{Cline:2024zhs}%
  \BibitemOpen
  \bibfield  {author} {\bibinfo {author} {\bibfnamefont {J.~M.}\ \bibnamefont
  {Cline}}\ }(\bibinfo {year} {2024})\ \Eprint
  {http://arxiv.org/abs/2401.02958} {arXiv:2401.02958 [hep-ph]} \BibitemShut
  {NoStop}%
\bibitem [{\citenamefont {Buccio}\ \emph {et~al.}(2024)\citenamefont {Buccio},
  \citenamefont {Donoghue}, \citenamefont {Menezes},\ and\ \citenamefont
  {Percacci}}]{Buccio:2024hys}%
  \BibitemOpen
  \bibfield  {author} {\bibinfo {author} {\bibfnamefont {D.}~\bibnamefont
  {Buccio}}, \bibinfo {author} {\bibfnamefont {J.~F.}\ \bibnamefont
  {Donoghue}}, \bibinfo {author} {\bibfnamefont {G.}~\bibnamefont {Menezes}}, \
  and\ \bibinfo {author} {\bibfnamefont {R.}~\bibnamefont {Percacci}},\ }\href
  {\doibase 10.1103/PhysRevLett.133.021604} {\bibfield  {journal} {\bibinfo
  {journal} {Phys. Rev. Lett.}\ }\textbf {\bibinfo {volume} {133}},\ \bibinfo
  {pages} {021604} (\bibinfo {year} {2024})},\ \Eprint
  {http://arxiv.org/abs/2403.02397} {arXiv:2403.02397 [hep-th]} \BibitemShut
  {NoStop}%
\bibitem [{\citenamefont {Stingl}(1986)}]{Stingl:1985hx}%
  \BibitemOpen
  \bibfield  {author} {\bibinfo {author} {\bibfnamefont {M.}~\bibnamefont
  {Stingl}},\ }\href {\doibase 10.1103/PhysRevD.36.651} {\bibfield  {journal}
  {\bibinfo  {journal} {Phys. Rev. D}\ }\textbf {\bibinfo {volume} {34}},\
  \bibinfo {pages} {3863} (\bibinfo {year} {1986})},\ \bibinfo {note}
  {[Erratum: Phys.Rev.D 36, 651 (1987)]}\BibitemShut {NoStop}%
\bibitem [{\citenamefont {Stingl}(1996)}]{Stingl:1994nk}%
  \BibitemOpen
  \bibfield  {author} {\bibinfo {author} {\bibfnamefont {M.}~\bibnamefont
  {Stingl}},\ }\href {\doibase 10.1007/BF01285154} {\bibfield  {journal}
  {\bibinfo  {journal} {Z. Phys. A}\ }\textbf {\bibinfo {volume} {353}},\
  \bibinfo {pages} {423} (\bibinfo {year} {1996})},\ \Eprint
  {http://arxiv.org/abs/hep-th/9502157} {arXiv:hep-th/9502157} \BibitemShut
  {NoStop}%
\bibitem [{\citenamefont {Alkofer}\ and\ \citenamefont {von
  Smekal}(2001)}]{Alkofer:2000wg}%
  \BibitemOpen
  \bibfield  {author} {\bibinfo {author} {\bibfnamefont {R.}~\bibnamefont
  {Alkofer}}\ and\ \bibinfo {author} {\bibfnamefont {L.}~\bibnamefont {von
  Smekal}},\ }\href {\doibase 10.1016/S0370-1573(01)00010-2} {\bibfield
  {journal} {\bibinfo  {journal} {Phys. Rept.}\ }\textbf {\bibinfo {volume}
  {353}},\ \bibinfo {pages} {281} (\bibinfo {year} {2001})},\ \Eprint
  {http://arxiv.org/abs/hep-ph/0007355} {arXiv:hep-ph/0007355 [hep-ph]}
  \BibitemShut {NoStop}%
\bibitem [{\citenamefont {Fischer}\ and\ \citenamefont
  {Alkofer}(2003)}]{Fischer:2003rp}%
  \BibitemOpen
  \bibfield  {author} {\bibinfo {author} {\bibfnamefont {C.~S.}\ \bibnamefont
  {Fischer}}\ and\ \bibinfo {author} {\bibfnamefont {R.}~\bibnamefont
  {Alkofer}},\ }\href {\doibase 10.1103/PhysRevD.67.094020} {\bibfield
  {journal} {\bibinfo  {journal} {Phys. Rev. D}\ }\textbf {\bibinfo {volume}
  {67}},\ \bibinfo {pages} {094020} (\bibinfo {year} {2003})},\ \Eprint
  {http://arxiv.org/abs/hep-ph/0301094} {arXiv:hep-ph/0301094} \BibitemShut
  {NoStop}%
\bibitem [{\citenamefont {Alkofer}\ \emph {et~al.}(2004)\citenamefont
  {Alkofer}, \citenamefont {Detmold}, \citenamefont {Fischer},\ and\
  \citenamefont {Maris}}]{Alkofer:2003jj}%
  \BibitemOpen
  \bibfield  {author} {\bibinfo {author} {\bibfnamefont {R.}~\bibnamefont
  {Alkofer}}, \bibinfo {author} {\bibfnamefont {W.}~\bibnamefont {Detmold}},
  \bibinfo {author} {\bibfnamefont {C.~S.}\ \bibnamefont {Fischer}}, \ and\
  \bibinfo {author} {\bibfnamefont {P.}~\bibnamefont {Maris}},\ }\href
  {\doibase 10.1103/PhysRevD.70.014014} {\bibfield  {journal} {\bibinfo
  {journal} {Phys. Rev. D}\ }\textbf {\bibinfo {volume} {70}},\ \bibinfo
  {pages} {014014} (\bibinfo {year} {2004})},\ \Eprint
  {http://arxiv.org/abs/hep-ph/0309077} {arXiv:hep-ph/0309077} \BibitemShut
  {NoStop}%
\bibitem [{\citenamefont {Cucchieri}\ \emph {et~al.}(2005)\citenamefont
  {Cucchieri}, \citenamefont {Mendes},\ and\ \citenamefont
  {Taurines}}]{Cucchieri:2004mf}%
  \BibitemOpen
  \bibfield  {author} {\bibinfo {author} {\bibfnamefont {A.}~\bibnamefont
  {Cucchieri}}, \bibinfo {author} {\bibfnamefont {T.}~\bibnamefont {Mendes}}, \
  and\ \bibinfo {author} {\bibfnamefont {A.~R.}\ \bibnamefont {Taurines}},\
  }\href {\doibase 10.1103/PhysRevD.71.051902} {\bibfield  {journal} {\bibinfo
  {journal} {Phys. Rev. D}\ }\textbf {\bibinfo {volume} {71}},\ \bibinfo
  {pages} {051902} (\bibinfo {year} {2005})},\ \Eprint
  {http://arxiv.org/abs/hep-lat/0406020} {arXiv:hep-lat/0406020} \BibitemShut
  {NoStop}%
\bibitem [{\citenamefont {Dudal}\ \emph {et~al.}(2008)\citenamefont {Dudal},
  \citenamefont {Gracey}, \citenamefont {Sorella}, \citenamefont
  {Vandersickel},\ and\ \citenamefont {Verschelde}}]{Dudal:2008sp}%
  \BibitemOpen
  \bibfield  {author} {\bibinfo {author} {\bibfnamefont {D.}~\bibnamefont
  {Dudal}}, \bibinfo {author} {\bibfnamefont {J.~A.}\ \bibnamefont {Gracey}},
  \bibinfo {author} {\bibfnamefont {S.~P.}\ \bibnamefont {Sorella}}, \bibinfo
  {author} {\bibfnamefont {N.}~\bibnamefont {Vandersickel}}, \ and\ \bibinfo
  {author} {\bibfnamefont {H.}~\bibnamefont {Verschelde}},\ }\href {\doibase
  10.1103/PhysRevD.78.065047} {\bibfield  {journal} {\bibinfo  {journal} {Phys.
  Rev. D}\ }\textbf {\bibinfo {volume} {78}},\ \bibinfo {pages} {065047}
  (\bibinfo {year} {2008})},\ \Eprint {http://arxiv.org/abs/0806.4348}
  {arXiv:0806.4348 [hep-th]} \BibitemShut {NoStop}%
\bibitem [{\citenamefont {Cucchieri}\ \emph {et~al.}(2012)\citenamefont
  {Cucchieri}, \citenamefont {Dudal}, \citenamefont {Mendes},\ and\
  \citenamefont {Vandersickel}}]{Cucchieri:2011ig}%
  \BibitemOpen
  \bibfield  {author} {\bibinfo {author} {\bibfnamefont {A.}~\bibnamefont
  {Cucchieri}}, \bibinfo {author} {\bibfnamefont {D.}~\bibnamefont {Dudal}},
  \bibinfo {author} {\bibfnamefont {T.}~\bibnamefont {Mendes}}, \ and\ \bibinfo
  {author} {\bibfnamefont {N.}~\bibnamefont {Vandersickel}},\ }\href {\doibase
  10.1103/PhysRevD.85.094513} {\bibfield  {journal} {\bibinfo  {journal} {Phys.
  Rev. D}\ }\textbf {\bibinfo {volume} {85}},\ \bibinfo {pages} {094513}
  (\bibinfo {year} {2012})},\ \Eprint {http://arxiv.org/abs/1111.2327}
  {arXiv:1111.2327 [hep-lat]} \BibitemShut {NoStop}%
\bibitem [{\citenamefont {Hayashi}\ and\ \citenamefont
  {Kondo}(2019)}]{Hayashi:2018giz}%
  \BibitemOpen
  \bibfield  {author} {\bibinfo {author} {\bibfnamefont {Y.}~\bibnamefont
  {Hayashi}}\ and\ \bibinfo {author} {\bibfnamefont {K.-I.}\ \bibnamefont
  {Kondo}},\ }\href {\doibase 10.1103/PhysRevD.99.074001} {\bibfield  {journal}
  {\bibinfo  {journal} {Phys. Rev. D}\ }\textbf {\bibinfo {volume} {99}},\
  \bibinfo {pages} {074001} (\bibinfo {year} {2019})},\ \Eprint
  {http://arxiv.org/abs/1812.03116} {arXiv:1812.03116 [hep-th]} \BibitemShut
  {NoStop}%
\bibitem [{\citenamefont {Binosi}\ and\ \citenamefont
  {Tripolt}(2020)}]{Binosi:2019ecz}%
  \BibitemOpen
  \bibfield  {author} {\bibinfo {author} {\bibfnamefont {D.}~\bibnamefont
  {Binosi}}\ and\ \bibinfo {author} {\bibfnamefont {R.-A.}\ \bibnamefont
  {Tripolt}},\ }\href {\doibase 10.1016/j.physletb.2019.135171} {\bibfield
  {journal} {\bibinfo  {journal} {Phys. Lett. B}\ }\textbf {\bibinfo {volume}
  {801}},\ \bibinfo {pages} {135171} (\bibinfo {year} {2020})},\ \Eprint
  {http://arxiv.org/abs/1904.08172} {arXiv:1904.08172 [hep-ph]} \BibitemShut
  {NoStop}%
\bibitem [{\citenamefont {Li}\ \emph {et~al.}(2020)\citenamefont {Li},
  \citenamefont {Lowdon}, \citenamefont {Oliveira},\ and\ \citenamefont
  {Silva}}]{Li:2019hyv}%
  \BibitemOpen
  \bibfield  {author} {\bibinfo {author} {\bibfnamefont {S.~W.}\ \bibnamefont
  {Li}}, \bibinfo {author} {\bibfnamefont {P.}~\bibnamefont {Lowdon}}, \bibinfo
  {author} {\bibfnamefont {O.}~\bibnamefont {Oliveira}}, \ and\ \bibinfo
  {author} {\bibfnamefont {P.~J.}\ \bibnamefont {Silva}},\ }\href {\doibase
  10.1016/j.physletb.2020.135329} {\bibfield  {journal} {\bibinfo  {journal}
  {Phys. Lett. B}\ }\textbf {\bibinfo {volume} {803}},\ \bibinfo {pages}
  {135329} (\bibinfo {year} {2020})},\ \Eprint
  {http://arxiv.org/abs/1907.10073} {arXiv:1907.10073 [hep-th]} \BibitemShut
  {NoStop}%
\bibitem [{\citenamefont {Fischer}\ and\ \citenamefont
  {Huber}(2020)}]{Fischer:2020xnb}%
  \BibitemOpen
  \bibfield  {author} {\bibinfo {author} {\bibfnamefont {C.~S.}\ \bibnamefont
  {Fischer}}\ and\ \bibinfo {author} {\bibfnamefont {M.~Q.}\ \bibnamefont
  {Huber}},\ }\href {\doibase 10.1103/PhysRevD.102.094005} {\bibfield
  {journal} {\bibinfo  {journal} {Phys. Rev. D}\ }\textbf {\bibinfo {volume}
  {102}},\ \bibinfo {pages} {094005} (\bibinfo {year} {2020})},\ \Eprint
  {http://arxiv.org/abs/2007.11505} {arXiv:2007.11505 [hep-ph]} \BibitemShut
  {NoStop}%
\bibitem [{\citenamefont {Huber}(2020)}]{Huber:2020keu}%
  \BibitemOpen
  \bibfield  {author} {\bibinfo {author} {\bibfnamefont {M.~Q.}\ \bibnamefont
  {Huber}},\ }\href {\doibase 10.1103/PhysRevD.101.114009} {\bibfield
  {journal} {\bibinfo  {journal} {Phys. Rev. D}\ }\textbf {\bibinfo {volume}
  {101}},\ \bibinfo {pages} {114009} (\bibinfo {year} {2020})},\ \Eprint
  {http://arxiv.org/abs/2003.13703} {arXiv:2003.13703 [hep-ph]} \BibitemShut
  {NoStop}%
\bibitem [{\citenamefont {Horak}\ \emph
  {et~al.}(2022{\natexlab{a}})\citenamefont {Horak}, \citenamefont
  {Pawlowski},\ and\ \citenamefont {Wink}}]{Horak:2022myj}%
  \BibitemOpen
  \bibfield  {author} {\bibinfo {author} {\bibfnamefont {J.}~\bibnamefont
  {Horak}}, \bibinfo {author} {\bibfnamefont {J.~M.}\ \bibnamefont
  {Pawlowski}}, \ and\ \bibinfo {author} {\bibfnamefont {N.}~\bibnamefont
  {Wink}},\ }\href@noop {} {\  (\bibinfo {year} {2022}{\natexlab{a}})},\
  \Eprint {http://arxiv.org/abs/2202.09333} {arXiv:2202.09333 [hep-th]}
  \BibitemShut {NoStop}%
\bibitem [{\citenamefont {Braun}\ \emph {et~al.}(2023)\citenamefont {Braun}
  \emph {et~al.}}]{Braun:2022mgx}%
  \BibitemOpen
  \bibfield  {author} {\bibinfo {author} {\bibfnamefont {J.}~\bibnamefont
  {Braun}} \emph {et~al.},\ }\href {\doibase 10.21468/SciPostPhysCore.6.3.061}
  {\bibfield  {journal} {\bibinfo  {journal} {SciPost Phys. Core}\ }\textbf
  {\bibinfo {volume} {6}},\ \bibinfo {pages} {061} (\bibinfo {year} {2023})},\
  \Eprint {http://arxiv.org/abs/2206.10232} {arXiv:2206.10232 [hep-th]}
  \BibitemShut {NoStop}%
\bibitem [{\citenamefont {Horak}\ \emph
  {et~al.}(2022{\natexlab{b}})\citenamefont {Horak}, \citenamefont {Pawlowski},
  \citenamefont {Rodr\'\i{}guez-Quintero}, \citenamefont {Turnwald},
  \citenamefont {Urban}, \citenamefont {Wink},\ and\ \citenamefont
  {Zafeiropoulos}}]{Horak:2021syv}%
  \BibitemOpen
  \bibfield  {author} {\bibinfo {author} {\bibfnamefont {J.}~\bibnamefont
  {Horak}}, \bibinfo {author} {\bibfnamefont {J.~M.}\ \bibnamefont
  {Pawlowski}}, \bibinfo {author} {\bibfnamefont {J.}~\bibnamefont
  {Rodr\'\i{}guez-Quintero}}, \bibinfo {author} {\bibfnamefont
  {J.}~\bibnamefont {Turnwald}}, \bibinfo {author} {\bibfnamefont {J.~M.}\
  \bibnamefont {Urban}}, \bibinfo {author} {\bibfnamefont {N.}~\bibnamefont
  {Wink}}, \ and\ \bibinfo {author} {\bibfnamefont {S.}~\bibnamefont
  {Zafeiropoulos}},\ }\href {\doibase 10.1103/PhysRevD.105.036014} {\bibfield
  {journal} {\bibinfo  {journal} {Phys. Rev. D}\ }\textbf {\bibinfo {volume}
  {105}},\ \bibinfo {pages} {036014} (\bibinfo {year} {2022}{\natexlab{b}})},\
  \Eprint {http://arxiv.org/abs/2107.13464} {arXiv:2107.13464 [hep-ph]}
  \BibitemShut {NoStop}%
\bibitem [{\citenamefont {Horak}\ \emph {et~al.}(2023)\citenamefont {Horak},
  \citenamefont {Pawlowski}, \citenamefont {Turnwald}, \citenamefont {Urban},
  \citenamefont {Wink},\ and\ \citenamefont {Zafeiropoulos}}]{Horak:2023xfb}%
  \BibitemOpen
  \bibfield  {author} {\bibinfo {author} {\bibfnamefont {J.}~\bibnamefont
  {Horak}}, \bibinfo {author} {\bibfnamefont {J.~M.}\ \bibnamefont
  {Pawlowski}}, \bibinfo {author} {\bibfnamefont {J.}~\bibnamefont {Turnwald}},
  \bibinfo {author} {\bibfnamefont {J.~M.}\ \bibnamefont {Urban}}, \bibinfo
  {author} {\bibfnamefont {N.}~\bibnamefont {Wink}}, \ and\ \bibinfo {author}
  {\bibfnamefont {S.}~\bibnamefont {Zafeiropoulos}},\ }\href {\doibase
  10.1103/PhysRevD.107.076019} {\bibfield  {journal} {\bibinfo  {journal}
  {Phys. Rev. D}\ }\textbf {\bibinfo {volume} {107}},\ \bibinfo {pages}
  {076019} (\bibinfo {year} {2023})},\ \Eprint
  {http://arxiv.org/abs/2301.07785} {arXiv:2301.07785 [hep-ph]} \BibitemShut
  {NoStop}%
\bibitem [{\citenamefont {Jakovac}\ \emph {et~al.}(2016)\citenamefont
  {Jakovac}, \citenamefont {Kaposvari},\ and\ \citenamefont
  {Patkos}}]{Jakovac:2015iqa}%
  \BibitemOpen
  \bibfield  {author} {\bibinfo {author} {\bibfnamefont {A.}~\bibnamefont
  {Jakovac}}, \bibinfo {author} {\bibfnamefont {I.}~\bibnamefont {Kaposvari}},
  \ and\ \bibinfo {author} {\bibfnamefont {A.}~\bibnamefont {Patkos}},\
  }\bibfield  {booktitle} {\emph {\bibinfo {booktitle} {{Proceedings, Gribov-85
  Memorial Workshop on Theoretical Physics of XXI Century: Chernogolovka,
  Russia, June 7-20, 2015}}},\ }\href {\doibase 10.1142/S0217751X16450421}
  {\bibfield  {journal} {\bibinfo  {journal} {Int. J. Mod. Phys.}\ }\textbf
  {\bibinfo {volume} {A31}},\ \bibinfo {pages} {1645042} (\bibinfo {year}
  {2016})},\ \Eprint {http://arxiv.org/abs/1510.05782} {arXiv:1510.05782
  [hep-th]} \BibitemShut {NoStop}%
\bibitem [{\citenamefont {Osterwalder}\ and\ \citenamefont
  {Schrader}(1973)}]{Osterwalder:1973dx}%
  \BibitemOpen
  \bibfield  {author} {\bibinfo {author} {\bibfnamefont {K.}~\bibnamefont
  {Osterwalder}}\ and\ \bibinfo {author} {\bibfnamefont {R.}~\bibnamefont
  {Schrader}},\ }\href {\doibase 10.1007/BF01645738} {\bibfield  {journal}
  {\bibinfo  {journal} {Commun. Math. Phys.}\ }\textbf {\bibinfo {volume}
  {31}},\ \bibinfo {pages} {83} (\bibinfo {year} {1973})}\BibitemShut {NoStop}%
\end{thebibliography}%


\end{document}